\documentclass[journal]{IEEEtran}

\makeatletter
\long\def\@makecaption#1#2{\ifx\@captype\@IEEEtablestring%
\footnotesize\begin{center}{\normalfont\footnotesize #1}\\
{\normalfont\footnotesize\scshape #2}\end{center}%
\@IEEEtablecaptionsepspace
\else
\@IEEEfigurecaptionsepspace
\setbox\@tempboxa\hbox{\normalfont\footnotesize {#1.}~~ #2}%
\ifdim \wd\@tempboxa >\hsize%
\setbox\@tempboxa\hbox{\normalfont\footnotesize {#1.}~~ }%
\parbox[t]{\hsize}{\normalfont\footnotesize \noindent\unhbox\@tempboxa#2}%
\else
\hbox to\hsize{\normalfont\footnotesize\hfil\box\@tempboxa\hfil}\fi\fi}

\usepackage[compatibility=false]{caption}
\DeclareCaptionFont{quackfont}{\fontfamily{ptm}\fontsize{7.5pt}{9pt}\selectfont}
\usepackage[labelformat=simple,font=quackfont]{subcaption}

\makeatother
\usepackage{lipsum}
\makeatletter
\let\MYcaption\@makecaption
\usepackage{cite}
\usepackage{graphicx}
\usepackage{subcaption}
\usepackage{float}
\usepackage{refstyle}
\usepackage{multicol, blindtext}
\usepackage{amsmath}
\usepackage{amsfonts}
\usepackage{adjustbox}
\newcommand{\norm}[1]{\left\lVert#1\right\rVert}

\usepackage[font=scriptsize]{caption}
\usepackage[font=footnotesize]{caption}
\usepackage{stfloats}
\usepackage{lmodern,bm}                
\usepackage[T1]{sansmath}
\usepackage{xcolor}
\usepackage{algorithm, algorithmic}
\usepackage{enumitem}
\usepackage{hyperref}
\begin{document}

\title{{Complexity-Scalable Neural Network Based MIMO Detection With Learnable Weight Scaling}}

\author{Abdullahi~Mohammad,~\IEEEmembership{Student Member,~IEEE,}
        Christos~Masouros,~\IEEEmembership{Senior Member,~IEEE,}
        and~Yiannis~Andreopoulos,~\IEEEmembership{Senior Member,~IEEE}
        
\thanks{All authors are with the Department
of Electronic and Electrical Engineering, University College London, London WC1E 7JEK, UK. (e-mail: abdullahi.mohammad.16@ucl.ac.uk; c.masouros@ucl.ac.uk; i.andreopoulos@ucl.ac.uk)}
\thanks{This work was supported by EPSRC under grants EP/S028455/1 and EP/R035342/1, and in part by the Petroleum Technology Development Fund (PTDF) Overseas Scholarship Scheme, Nigeria under the award PTDF/ED/PHD/MA/1003/16. This work has been presented in part at the IEEE International Conference on Communications ICC 2020, Dublin, Ireland.}}

\markboth{IEEE Transactions on Communications 2020, TO APPEAR}%
{Submitted paper}

\maketitle

\begin{abstract}
This paper introduces a framework for systematic complexity scaling of deep neural network {(DNN)} based MIMO detectors. The model uses a fraction of the DNN inputs {by scaling their values through weights that follow}  monotonically non-increasing functions. This allows for weight scaling \textit{across} and \textit{within} the different {DNN} layers in order to achieve { accuracy-vs.-complexity scalability during inference}. {In order to further improve the performance of our proposal, we introduce a sparsity-inducing regularization constraint in conjunction with trainable weight-scaling functions. In this way, the network learns to balance detection accuracy versus complexity while also  increasing robustness to changes in the activation patterns, leading to further improvement in the detection accuracy and BER performance} at the same inference complexity. Numerical results show that our approach is {10-fold and 100-fold} less complex {than classical approaches based on} semi-definite relaxation and ML detection, respectively.
\end{abstract}


\IEEEpeerreviewmaketitle

\section{Introduction}

\IEEEPARstart{T} {he} emergence of smartphones, tablets, and various new applications in recent years has led to the dramatic increase in mobile data traffic, especially mobile video traffic and other multimedia applications ~\cite{wang2014cellular}. The developments underpinning Fifth Generation (5G) networks aim to improve the network capacity, spectral efficiency, and latency in order to accommodate these bandwidth consuming applications and services. A key technology for 5G is massive MIMO systems, where {the base stations (BSs) of the cellular network} are equipped with tens, hundreds or thousands of antennas \cite{agiwal2016next}. This configuration further muddles the signal detection problem for {the uplink transmission} due to the computational complexity of the detector, which increases as the number of {BS antennas} grows \cite{lu2014overview}.\par

MIMO detectors have been extensively studied over the last two decades \cite{yao2002lattice, windpassinger2006performance} with the view to improving their detection accuracy and decreasing their complexities. {The Maximum Likelihood (ML) detector is known to be optimal, but with prohibitive complexity \cite{kailath2005mimo}. The sphere decoder (SD) provides near optimal performance, but its complexity still scales exponentially with the number of antennas \cite{kailath2005mimo}. Optimization-based detectors such as: semidefine relaxation (SDR) and approximate message passing (AMP) have been proven to achieve near optimal performance in some scenarios, but at the expense of high-degree polynomial complexity in the case of the former and divergence due to iterative characteristics for the later \cite{sidiropoulos2006semidefinite, hung2010improved}. Finally, linear detection methods such as: zero forcing (ZF), maximum ratio combining (MCR) and minimum mean squared error (MMSE) \cite{wubben2004near} have the least complexity. Similarly, ZF and MMSE show good performance characteristics for massive MIMO systems \cite{zhang2018performance}, albeit still having a gap in detection accuracy versus ML detection} {as  the number of receive antennas decreases.} 

With the explosive growth of data, computers are empowered with some ``intelligence'', i.e., the ability to learn from data without being explicitly programmed. This concept is known as ``machine learning''. Deep learning (DL) is the next evolution of machine learning and has already brought unprecedented performance boosting in many fields: robotics, e-commerce, computer vision, and natural language processing \cite{Goodfellow-et-al-2016}. It offers a new paradigm for data-driven learning in problems that cannot be expressed by tractable mathematical models, or are challenged by algorithmic complexity. It is against this background that some research groups in the wireless communications community have started leveraging on the expressive powers of DL to find solutions to some physical layer problems where analytic models cannot be derived. Alternatively, when it is possible to have an analytic optimization objective, the analytic expression is highly non-convex and very-high dimensional, such that conventional numerical optimization is computationally unfeasible \cite{simeone2018very, qin2019deep}. Some of the recent work on learning {on} the physical layer involve channel coding  \cite{gruber2017deep, nachmani2016learning} and end-to-end communications, and signal detection \cite{o2016learning, o2017introduction, dorner2018deep}. In these works, the entire communications system blocks are implemented as an auto-encoder for learning to reconstruct the transmitted symbols at the receiver end. All these are limited to a single antenna at the transmitter and receiver ends.\par

More recent works have involved MIMO detection through deep learning. One of the earliest attempts is the work of O'Shea \textit{et al.} \cite{OShea2017DeepLB}, who implemented unsupervised learning using an auto-encoder as an extension of end-to-end learning of previous attempts \cite{o2017introduction}. Channel equalization for the nonlinear channel using a {DNN} was proposed by Xu \textit{et al.} \cite{xu2018joint}, where two neural networks are jointly trained. The first is a convolutional neural network (CNN), which is trained to recover the transmitted symbols from nonlinear distortions and channel impairments. The second is a multilayer perceptron (MLP), also known as the fully connected neural network, and is used to perform the detection.\par 
{The growing popularity of unfolding iterative optimization algorithms through projected gradient descent (deep-unfolding) to design DNNs to solve a spectrum of applications has led to a paradigm shift for efficient learning-based solutions for the physical layer design \cite{9020494}. One of the successful applications of deep-unfolding for MIMO detection is the ``DetNet" proposed by Samuel \textit{et al.} \cite{samuel2017deep}. The approach is significant as it derives a learnable signal detection architecture for multiple channels on a single training shot with near-optimal performance and also works well under both constant and Rayleigh fading channels. Multilevel MIMO detection using coupled-neural networks structure is investigated by Corlay \textit{et al.} \cite{corlay2018multilevel}. The network uses a multi-stage sigmoid activation function and a random forest approach to reduce the detection complexity with relatively fewer parameters. A similar approach by unfolding belief propagation (BP) based on modified BP algorithms (damped BP and maximum BP) is later introduced by Tan \textit{et al.} \cite{Tan2018ImprovingMM} and Liu and Li \cite{liu2018deep}. The work proposed in \cite{samuel2017deep} has been further extended to handle higher digital constellations \cite{samuel2019learning}, where the authors investigate the complexity-accuracy trade-off as more layers are added.}\par

{Beyond using projected gradient descent approaches with DNN architectures, other lower-cost learning-based detectors based on iterative AMP algorithms are: ``trainable iterative detector (TI-detector)" proposed by Imanishi \textit{et al.} \cite{Imanishi2018DeepLI}, ``orthogonal approximate massage passing deep network (OAMP-Net)" introduced by He \textit{et al.}\cite{he2018model} and ``trainable projected gradient detector (TPG-Net)” proposed by Takabe \textit{et al.} \cite{takabe2019trainable}. Contrary to previous learning-based detectors that heavily depend on a huge amount of parameters, these models exploit full domain knowledge to achieve acceptable performance with fewer parameters. However, these algorithms require channel inversion at every training and inference steps to compute the nonlinear estimator for symbol detection. On the other hand, the proposed weight-scaling neural-network (WeSNet) is a general framework for reducing the complexity of broader DNN-based receivers and therefore extends to numerous relevant neural network (NN) designs that do not embed NN architecture. Furthermore, WeSNet, as it applies to the DetNet unlike OAMP-Net and TPG-Net, does not require channel inversion, though the authors in \cite{takabe2019trainable} argue that for TPG-Net, matrix inversion is only required during the initialization when the channel is assumed constant}.\par 

{Generally, designing deeper NN architectures for signal detection problems comes with significantly increased training and inference complexity, while gains in detection performance are not always significantly increased.} This creates the imperative for systematic approaches to design DNN architectures with scalable complexity {that can speed up offline training (learning),\footnote{{In practice, training a deep neural network (DNN) is done offline and is computationally expensive in addition to requiring large training data. Generally, the performance of a trained DNN model is determined by its ability to generalize well on a new set of data (test data). Therefore, model testing is done online using Monte Carlo simulation with new channel instantiations at different SNR conditions at the edge of the device to evaluate the efficacy of the trained model.}} facilitate model deployment and inference on a range of devices such as mobile devices, and other embedded hardware platforms with limited resources.} It is against this background that DNN acceleration at both training and inference becomes an active research area within the deep learning community and has recently received significant attention \cite{sze2017efficient, he2017channel, zhu2018adaptive}. {Popular techniques of complexity reduction that are similar to our proposal in style are Dropout \cite{srivastava2014dropout}, Drop-Connect \cite{wan2013regularization} and Pruning \cite{he2017channel}. However, these schemes fundamentally differ from our proposal because most of them are used to prevent overfitting, and they are not explicitly designed (of applicable) for complexity reduction. The simplest of them is Dropout \footnote{{Dropout requires additional matrices for dropout masks, random selection of numbers for each entry of these matrices and matrix multiplication of the masks with the corresponding weights. At inference time, which is the focus of our work, Dropout uses the full network and does not allow for scalable complexity-accuracy adjustments.}} where some units (neurons) are randomly shot during training. At inference time however, Dropout uses the full network whereas our proposed framework allows for the network to dynamically adjust its computational complexity and detection accuracy characteristics at inference.} While many proposals have been put forward for accelerated DNN training and inference in computer vision \cite{rastegari2016xnor, wan2013regularization, hubara2016binarized, zhu2018adaptive}, to the best of our knowledge, no systematic DNN acceleration has so far been designed for physical layer communications. In this work, we attempt to fill this gap by proposing a complexity-scalable DNN model for efficient MIMO detection. \par

{Deep learning-based channel decoding is proposed as a solution to address the \textit{curse of dimensionality} \cite{wang1996artificial}. For example, as pointed out by Gruber \textit{et al.} \cite{gruber2017deep}, code length of 100 or larger and rate 0.5 requires $M\geq2^{50}$ different code-words, and it is extremely difficult to fully train a generic NN architecture to decode a message. However, this could be addressed by allowing the network to learn some decoding strategies from the structured codewords to enable it to derive inference over the entire code-book \cite{wang1996artificial}}. 

In this work, we introduce the concept of monotonic non-increasing profile function that scale {each layer of the NN} in order to allow the network to dynamically learn the best attenuation strategy for its own weights during training. {By doing so, we introduce sparsity in the DNN, which results in a significant complexity saving at inference}. Our focus is on DNN designs that unfold iterative projected gradient descent unconstrained optimization for massive MIMO ML-based detection. While methods of artificial ``suppression'' of neurons during training are known to create sparsity and can be detrimental to inference accuracy  \cite{he2017channel,mcdanel2017incomplete}, we show that, by tuning these profile function appropriately, we can provide a control mechanism that trades off DNN complexity for detection accuracy in a scalable manner.
{Our contributions are  summarized below:}
{\begin{itemize}
    \item {We introduce} a weight scaling framework for DNN-based MIMO detection. Our approach is realized by adjusting layer weights through monotonic profile functions. The original DNN design is based on DetNet, i.e., unfolding a projected gradient descent scheme \cite{samuel2017deep}. We term our proposal the weight-scaling neural-network based MIMO detector (WeSNet). 
    \item In order to allow for entire layers to be abrogated in a controllable manner during inference, we introduce a regularization approach that imposes constraints on the layer weights. {This allows for scalable reduction in the model size and the incurred computational complexity, with graceful degradation in the detection accuracy}. 
    \item To improve the performance of WeSNet, we introduce a learnable accuracy-complexity design, where the weight profile functions themselves are made trainable in order to prevent vanishing gradients due to changes in the values of activations. This improves the detection accuracy of the WeSNet at the cost of increased memory due to increase in the model parameters.
    \item Finally, we present a comprehensive complexity analysis of  WeSNet inference in relation to learning-based MIMO detector(DetNet) and traditional detectors. Our study and results show that under the same experimental conditions,  WeSNet with 50\% of the layer weights outperforms the detection accuracy of DetNet while offering 51.43\% reduction in complexity and close to 50\% reduction in model size. Furthermore, its detection accuracy is similar to SDR with nearly 10-fold reduction of computational complexity.
\end{itemize}}

The remainder of the paper is structured as follows: System model and the review of traditional MIMO detectors are presented in Section \ref{section2}. The proposed approach is introduced in Section \ref{section3} and extensions are proposed in Section \ref{R-WeSNet}. Simulations and complexity results are presented in Section \ref{section4}. Finally, Section \ref{section5} summarizes and concludes the paper.\par

{\textbf{Notations:} We use bold lowercase symbols for vectors, lowercase symbols for scalars and bold uppercase symbols for matrices. Symbols $\Phi\{\mathord{\cdot}\}$ and $\tilde{\Phi}\{\mathord{\cdot}\}$ are used to denote the sets of trainable parameters. The auxiliary input vector is denoted by $\mathbf{\hat{a}}_{r}$.}   

\section{System Model and Review of MIMO Detectors}\label{section2}
Consider a communications system with \(N_t\) transmit and \(N_r\) receive antennas. The received signal is modelled using a standard MIMO channels equation as 
{\begin{equation}
\label{eq:equation1}
\bar{\mathbf{y}}=\bar{\mathbf{H}}{\bar{\bm{s}}} + \bar{\bm{n}}
\end{equation}
where the received complex symbol vector is: $ \bar{\mathbf{y}}\in \mathbf{\mathfrak{C}}^{N_{r}\times1}$, and the corresponding transmitted symbol vector, is $ \bar{\bm{s}}\in \mathbf{\mathfrak{C}}^{N_{t}\times1}$. $\bar{\mathbf{H}}\in \mathbf{\mathfrak{C}}^{N_{r}\times N_{t}}$ is Rayleigh fading channel matrix and $\bar{\bm{n}}\in \mathbb{\mathfrak{C}}^{N_{r}\times1}$ is the additive white Gaussian noise (AWGN) vector with zero mean and variance $\displaystyle \sigma ^{2}$. For convenience and ease of implementation, the channel model is redefined over the real domain as
\begin{align*}
\mathbf{y} \equiv \begin{bmatrix}
\Re \{\bar{\mathbf{{y}}}\}\\
\Im \{\bar{\mathbf{y}}\}
\end{bmatrix} \in \mathbb{R}^{2N_{r}\times1} ,\ \bm{s} \equiv \begin{bmatrix}
\Re \{\bar{\bm{s}}\}\\
\Im \{\bar{\bm{s}}\}
\end{bmatrix} \in \mathbb{R}^{2N_{t}\times1}
\end{align*} 
\begin{align}
\mathbf{H} \equiv \begin{bmatrix}
\Re \{\bar{\mathbf{H}}\} & -\Im \{\bar{\mathbf{H}}\}\\
\Im \{\bar{\mathbf{H}}\} & \Re \{\bar{\mathbf{H}}\}
\end{bmatrix} \in \mathbb{R}^{2N_{r} \times 2N_{t}}
\end{align}} 

With this representation, (\ref{eq:equation1}) can be expressed in terms of real-valued vectors and matrix as
\begin{equation}
    \mathbf{y}=\mathbf{H}{\bm{s}} + \bm{n}
    \label{eq:system_model}
\end{equation}
{Below, we summarize few MIMO detectors that form our performance benchmarks.} 

\subsection{{Maximum} Likelihood-detector (ML-detector)}
The ML detector is known as the optimal detector \cite{ma2004semidefinite}. However, finding the estimated symbols involves searching over all the possible transmitted digital symbols of $|\mathbb{M}|^{N_{t}}$ vectors (where $\mathbb{M}$ is the constellation set and $N_{t}$ is number of transmit antennas). Therefore, the complexity of this detector grows exponentially with the number of transmit antennas and the modulation order (constellation size), and quickly becomes impractical in real systems. {ML-detector minimizes the Euclidean distance between the received and the transmitted symbols \cite{ma2004semidefinite}.} 
\begin{equation}
\label{eq:equation2}
\bm{\hat{s}}=\operatorname*{arg\,min}_{\bm{s}\in \bm{S}} \norm{\mathbf{y}-\mathbf{H}\bm{s}}^2
\end{equation}
{where $\bm{S}$ is the constellation set defined by the modulation scheme used (BPSK and 4-QAM)}

\subsection{Optimization Based Detector}
{Typical optimization based detectors are based on quadratically constrained quadratic programming (QCQP) to provide detection at lower computational cost than an ML detector \cite{5447068}. \Eqref{equation2} can be written as \cite{wiesel2005semidefinite}
\begin{equation}
\begin{aligned}
& \underset{\bm{X}}{\text{min}}
& & \mathrm{trace}(\mathbf{LX}) \\
& \text{s.t.}
& & \text{diag}{(\mathbf{X})}=\mathbf{I} \\
&&& \mathbf{X}(2{N_t}+1,2{N_t}+1)=1 \\
&&& \mathbf{X} \succeq 0 \\
\label{eq:SDR3}
\end{aligned}
\end{equation}
where: 
\begin{align*}
\mathbf{L} =\ \begin{bmatrix}
\mathbf{H}^{T}\mathbf{H} & -\mathbf{H}^{T}\mathbf{y}\\
-\mathbf{y^{T} H} & \| \mathbf{y} \| ^{2}
\end{bmatrix}; \ \ \ \ \ & \mathbf{X} =\bm{s}^T\bm{s}    
\end{align*}
}

{\subsection{Deep MIMO Detectors}
Most relevant to  our work are detectors that use a {DNN} architecture  \cite{samuel2017deep,samuel2019learning}. The unique property of such detectors is their ability to sustain their performance under higher-dimensional signals \cite{samuel2019learning}. The premise of the operation of all such learning-based detectors is that the estimate of the received symbols is obtained from a trained network by an update rule using an iterative projected gradient descent formulation \cite{samuel2019learning}. For a function defined by $ f(x,y)$, the estimate of $\mathbf{x}$ and $\mathbf{y}$ over the $r$-{th} iteration (i.e., layer) can be found from gradient descent using the following update rule:
\begin{subequations}
\begin{align}
\mathbf{x}_{r+1}=\mathbf{x}_{r}-\eta\frac{\partial f(x,y)}{\partial x} \\
\mathbf{y}_{r+1}=\mathbf{y}_{r}-\eta\frac{\partial f(x,y)}{\partial y}
\end{align}
\end{subequations}
where $\eta$ is the learning rate}.\par 

{DetNet is designed by applying gradient descent optimization in (\ref{eq:equation2}) expressed as
\begin{equation}\label{eq:equationg1}
    \left.\hat{\bm{s}}_{r+1}=\hat{\bm{s}}_{r}-\mathbf{\eta}_r\frac{\partial \norm{\mathbf{y}-\mathbf{H}\bm{s}}^2}{\partial \bm{s}}\right\vert_{\bm{s}=\hat{\bm{s}}_{r}}
\end{equation}
By using $\mathbf{H}^T \mathbf{y} ,\ \mathbf{H}^T \mathbf{H}$ and $\bm{s}$ as inputs, and via the application of non-linear functions prior to the outputs, the formulation of (\ref{eq:equationg1}) is converted to three sublayers with each sublayer comprising a { perceptron}, also known as fully-connected NN. This is defined by the following equations}

{\begin{equation}
  \label{eq:equation10}
   \mathbf{u}_{r}=\Theta{{(\mathbf{W}_{1r}\mathbf{x}_{r}+\bm{b}_{1r})}}
\end{equation}}
where:

{\begin{equation}
  \mathbf{x}_{r} = \Pi(\mathbf{H}^T\mathbf{y},\ \mathbf{H}^T\mathbf{H}\bm{s}_{r},\ \bm{s}_{r},\ \mathbf{a}_{r})
\end{equation}}

{\begin{equation}
\label{eq:equation12}
\bm{\hat{s}}_{r+1}=\Psi{(\mathbf{W}_{2r}\mathbf{u}_{r}+\bm{b}_{2r})}  
\end{equation}
\begin{equation}
\label{eq:equation13}
\mathbf{\hat{a}}_{r+1}=\mathbf{W}_{3r}\mathbf{u}_{r}+\bm{b}_{3r} 
\end{equation}}
{{$\Pi(\mathord{\cdot})$ is the concatenation function, and $\Theta(\mathord{\cdot})$ and $\Psi(\mathord{\cdot})$ are nonlinear and piece-wise linear sign functions, respectively, and subscripts $1r$, $2r$ and $3r$ indicate the three sublayers of layer $r$}.
The trainable parameters that are optimized during training are defined by}

{\begin{equation}
\label{eq:trainable_params}
\mathbf{\Phi} =\{\mathbf{W}_{1r},\ \mathbf{W}_{2r} \ ,\ \mathbf{W}_{3r},\ \bm{b}_{1r},\ \bm{b}_{2r},\ \bm{b}_{3r} \}^{L}_{r=1}
\end{equation}}

\section{Proposed {Weight-Scaling} Neural-Network based MIMO Detector \normalfont{(WeSNet)}}\label{section3}

\subsection{Weight Scaling Vector Coefficient (WSVC)}
{In this section, we propose a scalable accuracy-complexity framework for DNN-based MIMO receivers through systematic weight scaling with monotonic non-increasing functions for both feed-forward and edge inference computations. This allows for our proposal to have minimum deployment friction as it allows for the best operating point in the accuracy-complexity sense to be devised at inference.} \par
\begin{figure}[!t]
    \centering
    \includegraphics[width=3.7in,height=2.7in]{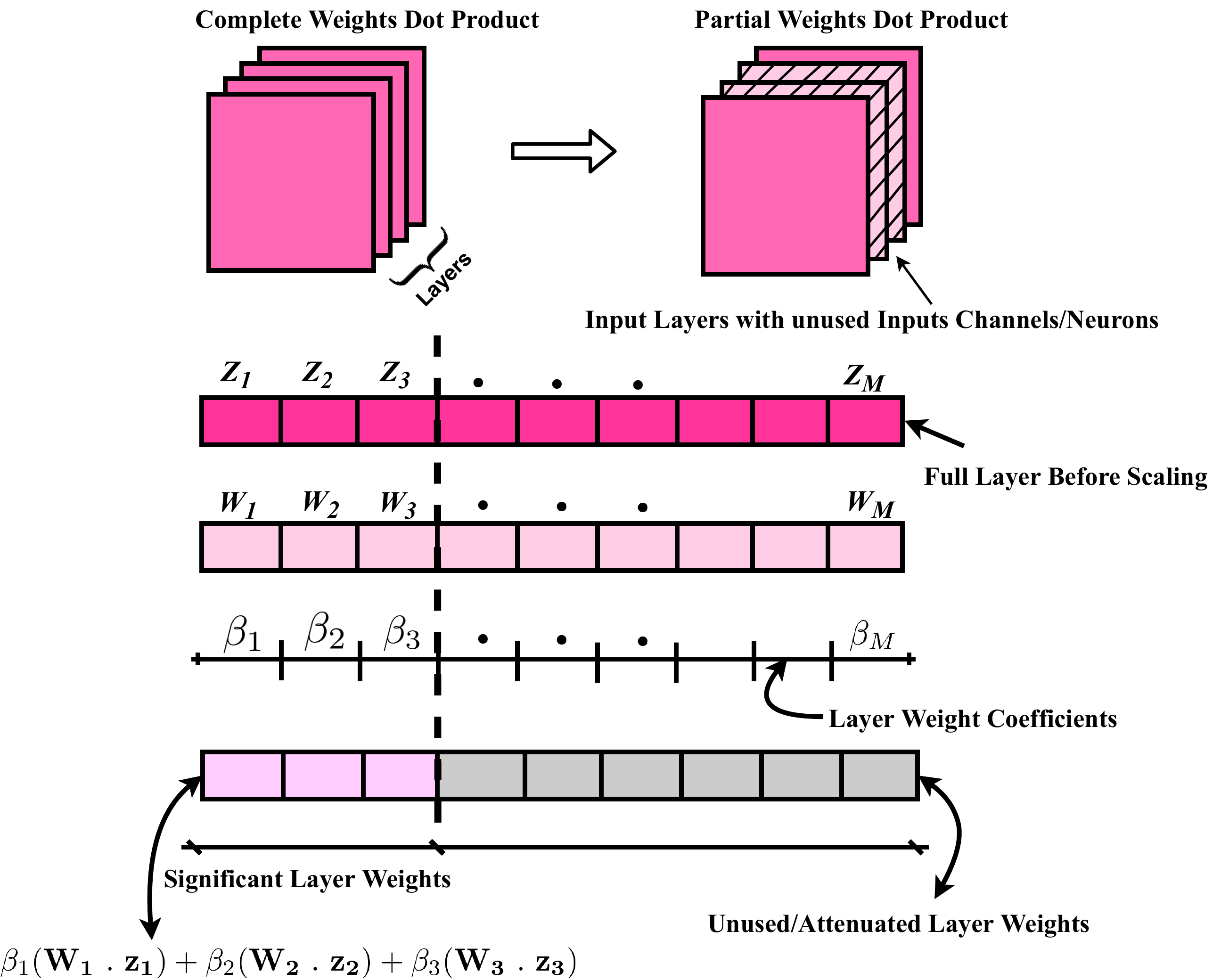}
    \DeclareGraphicsExtensions.
    \caption{{WSVC in a single layer of an MLP allowing for attenuated layer weights to (optionally) be dropped.}}
    \label{fig:WSVC}
\end{figure}

 
A WSVC is computed by applying monotonically non-increasing coefficients (known as profile function coefficients) to the layer weights during the forward propagation. This results in prioritizing the selection of the layer weights in decreasing fashion from the {most significant to least significant}. Mathematically, for two given vectors, $\mathbf{x} =[ x_{1} ,x_{2} ,\ .\ .\ .\ ,x_{N}]^{T}$ and $\mathbf{y}=[ y_{1} ,y_{2} ,\ .\ .\ .\ ,y_{N}]^T$, if $\bm{\beta} $ is the vector of the profile coefficients, WSVC is the pruned version of the form
\begin{equation}
    \sum_{i=1}^{N}{\beta}_{i}{{x}_{i}}{{y}_{i}} = \beta_{1}{{x}_{1}}{{y}_{1}}+\beta_{2}{{x}_{2}}{{y}_{2}}, . . .+\beta_{N}{{x}_{N}}{{y}_{N}}
    \end{equation}
In a standard fully connected NN, the output of the feed forward pass is given by
\begin{equation}
    {z}_{j}=\sum_{i=1}^{N}{{W}_{ji}}{{x}_{i}}+{b}_{j} 
\end{equation}
where $i$ and $j$ are the input and output dimensions (size of the neurons) respectively; ${x}_{i}$ is the $i$-th input components, ${W}_{ji}$ is the channel or layer weight corresponding to the $j$th output and ${b}_{j}$ is the output bias. The corresponding WSVC is derived by
\begin{equation}
    {z}_{j}=\sum_{i=1}^{N}\beta_{i}{{W}_{ji}}{{x}_{i}}+{b}_{j} 
\end{equation}
 Fig. \ref{fig:WSVC} shows the difference between the feed forward computations of a layer of an MLP and the MLP augmented by WSVC. The part of the WSVC corresponding to significant layer weights is indicated by the light colored shaded region on the bottom-right side of the figure. The example shows that, via the WSVC, we can compute and use only one-third of the channel/layer weights out of the $N$ layer dimension, as the remaining two-thirds of the weights are attenuated and can be dropped. 
\subsection{Weight Coefficient Profile Function}
\begin{figure}[!t]
    \centering
    \includegraphics[width=\linewidth]{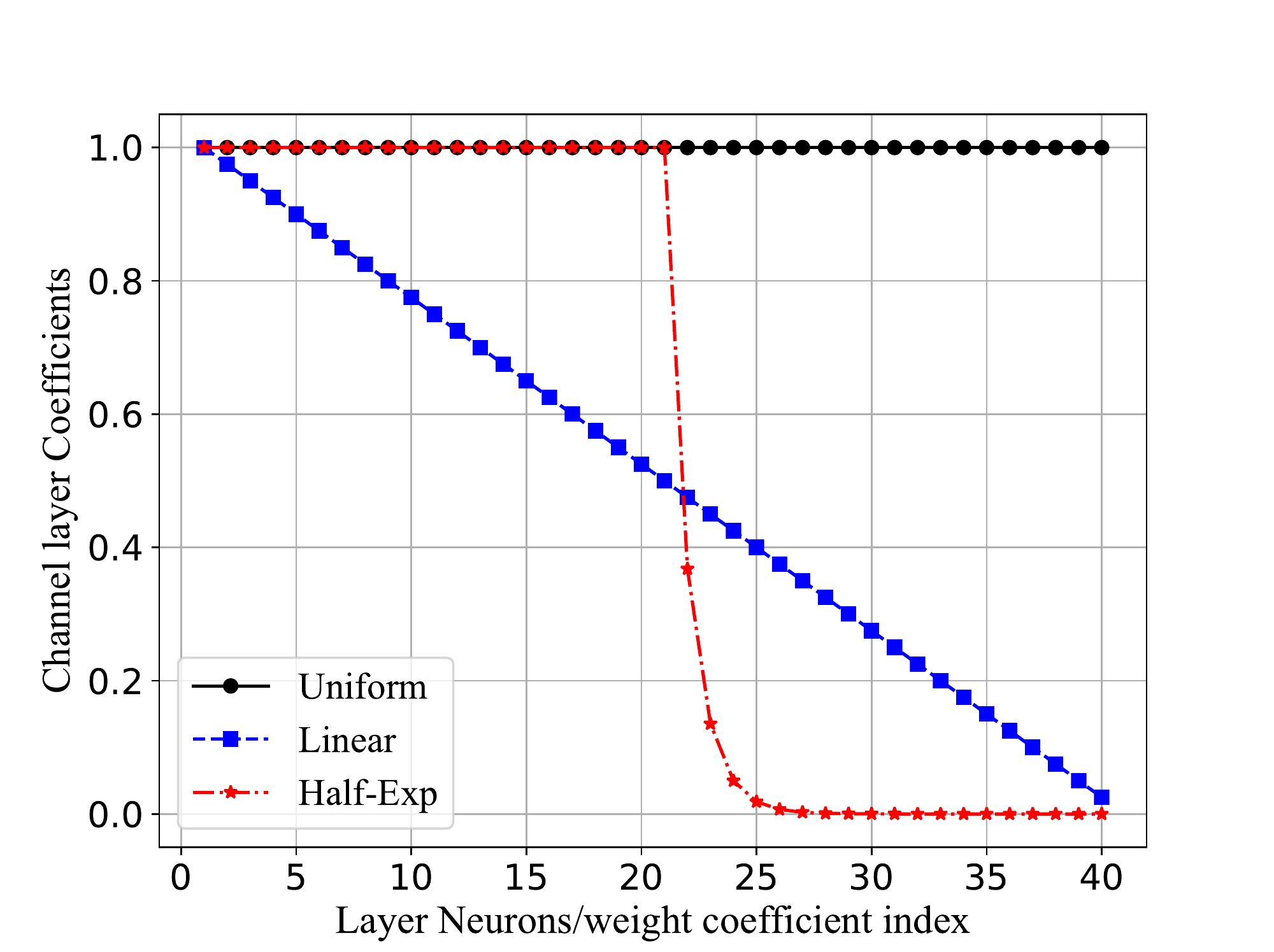}
    \DeclareGraphicsExtensions.
    \caption{Profile coefficients vs Neurons/Layer weight index.} 
    \label{fig:profile}
\end{figure}

We begin by introducing two non-increasing monotonic profile functions (Linear and Half-Exponential functions) for the weight coefficients \cite{mcdanel2017incomplete} as shown in Fig. \ref{fig:profile}.

\subsubsection{Linear Profile Function}
{This function comprises the profile coefficients obtained from the linear equation of the form:}
\begin{equation}
\label{eq:linear_prof}
\beta _{i} =1-\frac{i}{N} \ ;\ \forall \ i=1,\ 2,\ .\ .\ .\ ,\ N
\end{equation}
where $N$ is the layer size. 
\subsubsection{Half-Exponential Profile Function}
{This is a hybrid profile function from uniform and exponential functions.} This function attenuates coefficients corresponding to half of the channel via an exponential decay function. The implication of this is that it allows the network training to adjust the gradient flow such that important weights are retained in the non-attenuated half of each layer and the less important ones in the exponentially-attenuated half.
\begin{equation}
\label{eq:exp_prof}
\beta _{i} =\begin{cases}
1 & \ if\ i\ \leq \frac{N}{2} \ \forall \ i\ =\ 1,\ 2,\ .\ .\ .\ ,N \\
exp\left(\frac{N}{2} -i-1\right) & otherwise
\end{cases}
\end{equation}

\subsection{Structure of the WeSNet-Detector}
WeSNet is a nonlinear estimator designed by unfolding the ML metric using a recursive formulation of the projected gradient descent optimization. Our proposed detector applies the profile coefficients on the existing DetNet. Such a modification reduces the computational complexity for training the detector. {We apply profile coefficients to (\ref{eq:equation10}) and (\ref{eq:equation13}) to obtain the following non-linear WSVCs over the $i$-{th} and $j$-{th} inputs of the first and third sublayers of the $r$-{th} layer, respectively.}

\begin{figure}[!t]
\begin{subfigure}{.5\textwidth}
    \centering
    \includegraphics[width=\linewidth]{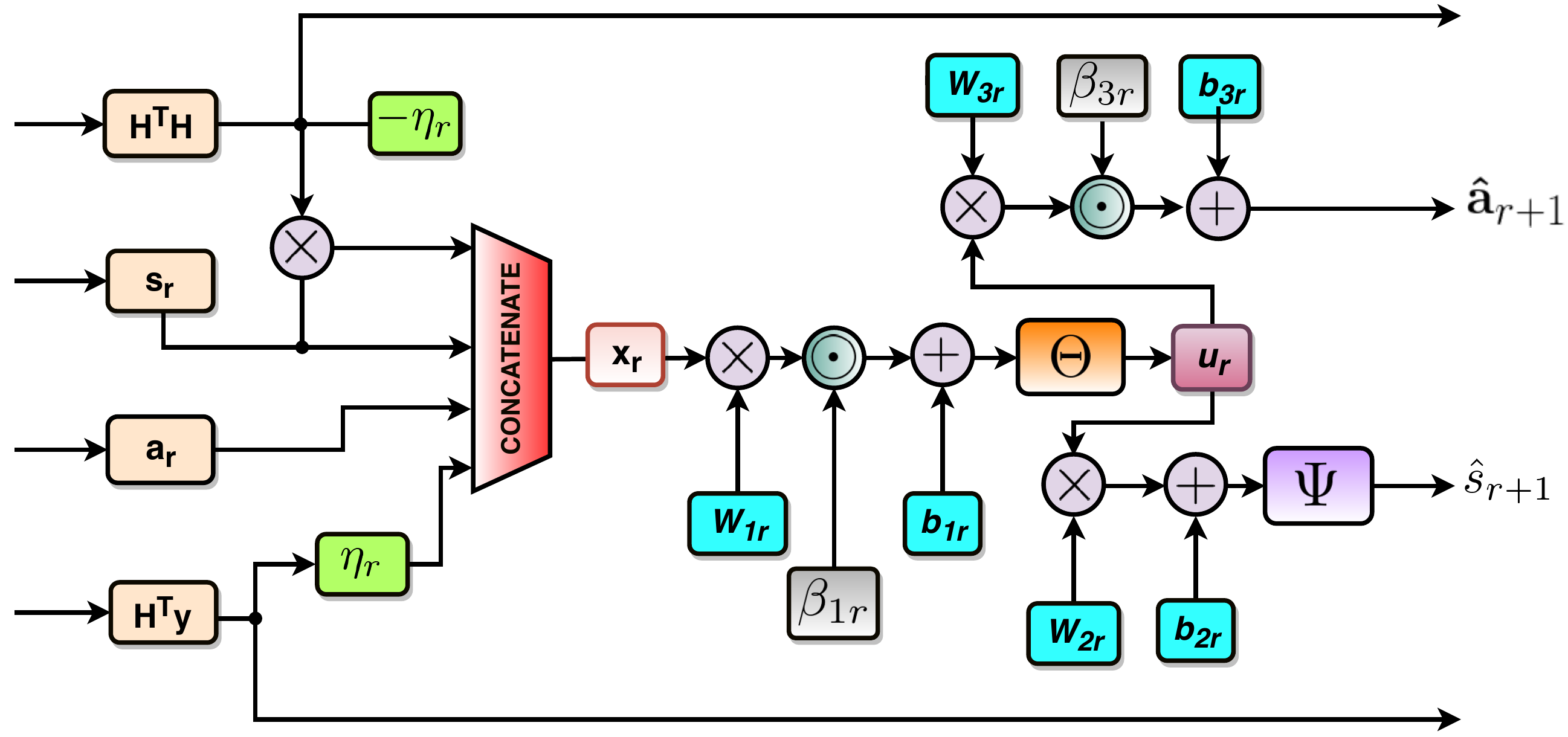}
    \DeclareGraphicsExtensions.
    \caption{Single $r$-{th} layer WeSNet-detector.}
    \label{fig:WSVC_DEC1}
    \vspace{4mm}
    \end{subfigure}
    \begin{subfigure}{.5\textwidth}
    \centering
    \includegraphics[width=\linewidth]{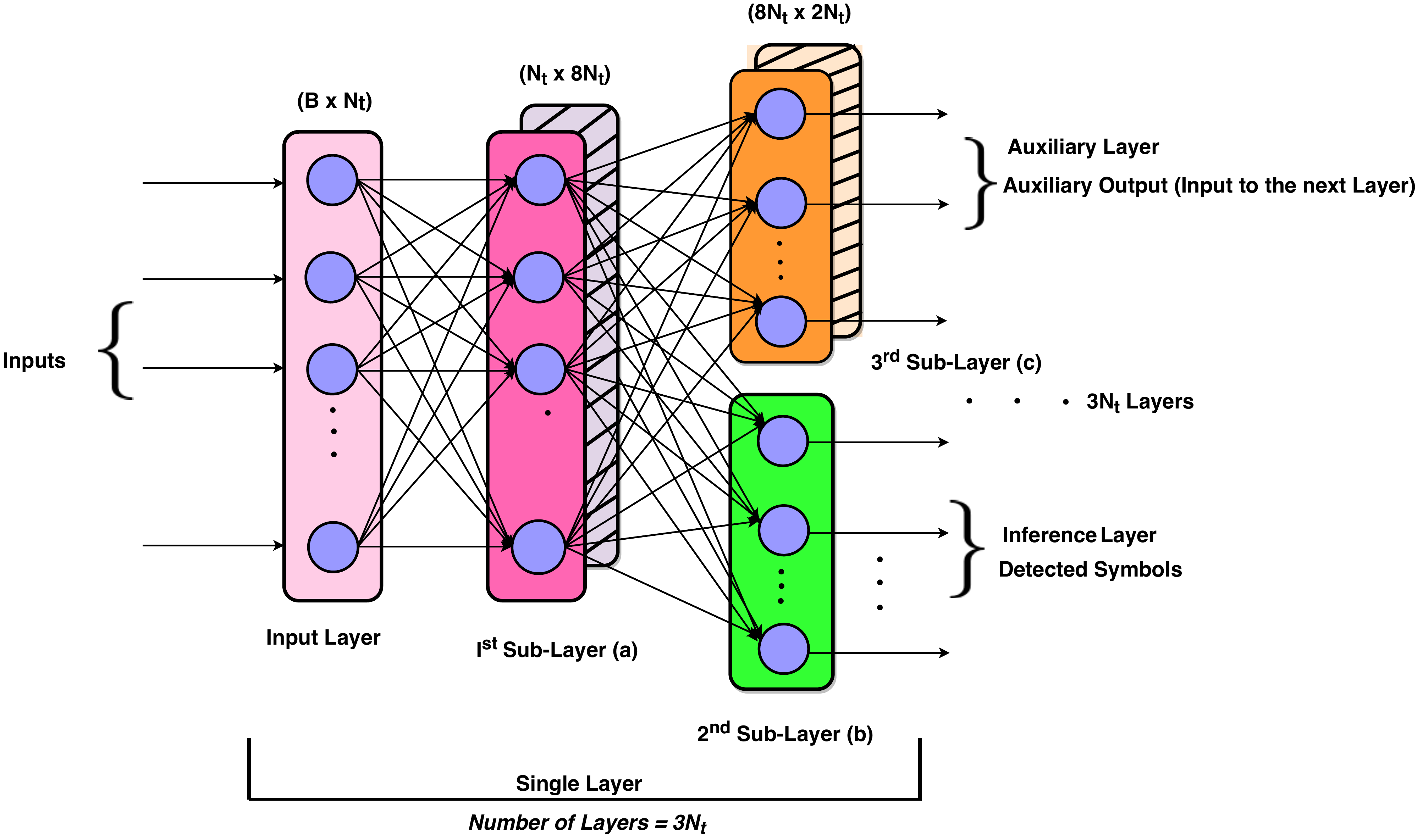}
    \DeclareGraphicsExtensions.
    \caption{Single Layer WeSNet Architecture.}
    \label{fig:WSVC_model}
    \end{subfigure}
\caption{WeSNet Model Architecture}
\label{fig:WeSNet_Model_Arch}
\end{figure}
{\begin{equation}
  \label{eq:prof1}
   u_{j[r]}=\Theta\left \{\sum_{i=1}^{N}{{\beta_{i[1r]}W_{ji[1r]}x_{i[r]}}}+{b}_{j[1r]}\right \}
\end{equation}}
{\begin{equation}
  \label{eq:prof2}
  {\hat{\text{a}}_{k[r+1]}=\sum_{j=1}^{M}\beta_{j[3r]}W_{kj[3r]}u_{j[r]}}+{b}_{k[3r]} 
\end{equation}}
{where: $j$ and $k$ are the outputs of the first and third sublayers of layers $r$ and $r+1$ respectively,  $N$ and $M$ are their corresponding sizes, and bracketed subscripts are added to explicitly indicate the membership of components to their corresponding network layers and sublayers.}\par
WeSNet has $3N_{t}$ layers with each layer having three sub-layers, the input layer, the auxiliary and the detection layer. The layer weights of the last sub-layer (detection layer) described by (\ref{eq:equation12}) are not scaled in order to maintain the full dimension of the detected symbols as originally transmitted. The flowchart of a single-layer WeSNet based on the (\ref{eq:equation12}), (\ref{eq:prof1}) and (\ref{eq:prof2}) is shown in Fig. \ref{fig:WSVC_DEC1}. The complete architecture of the WeSNet is shown in Fig. \ref{fig:WSVC_model}. Since the error estimation of the ML-detector does not require the knowledge of the noise variance, the loss function of WeSNet is derived as the weighted sum of the detector's errors normalized with the loss function of the standard linear inverse detector (ZF)\cite{samuel2017deep} as

\begin{equation}
\label{eq:loss_func}
    \mathcal{L}(s;\hat{s}(H,y:\Phi))=\sum_{r=1}^{L}\log({r})\frac{{\norm{\bm{s}-\bm{\hat{s}}_{r}}^2}}{{\norm{\bm{s}-{\tilde{\bm{s}}}}^2}}
\end{equation}

\section{Introducing Robustness through Regularized \normalfont{WeSNet (R-WeSNet)}}\label{R-WeSNet}
{In this section, we introduce log-regularization with a sparsity-enforcing mechanism. Unlike other proposals that employ such mechanisms as the means to avoiding over-fitting, the combination of our log-regularization with the proposed profile functions enables the network to learn the best profile function scaling to gracefully trade-off accuracy and complexity. Importantly, this achieves \textit{scalable} accuracy-complexity operation at inference by simply discarding parts of network layers (or even entire layers).}

\subsection{Motivation}
{Given that our aim is to introduce sparsity in conjunction with our profile function coefficients} so that layers (and parts of layers) with few non-zero coefficients can be removed to scale complexity, we propose the use of a {\textit{log}-$L_{1}$ norm.} The choice of $L_{1}$ is motivated by the fact that it forces some of the coefficients to be zero and leads to sparsity \cite{ scardapane2017group}, thereby making it more appealing and robust than $L_{2}$, as well as a better candidate for feature selection.

\subsection{Proposed Loss Function}
Following this motivation, the loss function of (\ref{eq:loss_func}) is modified {such that a \textit{log}-$L_{1}$ penalty term is imposed on the weights}:

\begin{equation}
\label{eq:reg_loss_func}
    \mathcal{L}(s;\hat{s}(H,y:\Phi))=\sum_{r=1}^{L}\log({r})\frac{{\norm{\bm{s}-\hat{\bm{s}}_{r}}^2}}{{\norm{\bm{s}-{\tilde{\bm{s}}}}^2}} + \lambda{f({\beta}_{r},{\tilde{W}}_{r})}
\end{equation}
where $\lambda$ is the regularization parameter that controls the importance of sparsity in the layers weights and $f({\beta}_{r},{\tilde{W}}_{r})$ is the function of layer weights with respect to the neuron connections between adjacent layers, and is given by
\begin{equation}
    \label{eq:regularization_func}
    f({\beta}_{r},{\tilde{W}}_{r})=\sum_{r=l}^{L}{\log{(1+(r-1){|{\beta}_{r}{{\tilde{W}}}_{r}|})}} \ \forall\ r = l, .\ . \ .\ , L
\end{equation}
where: 
\begin{equation}
  \left({\beta}{\tilde{W}}\right)(r,k)=\sum_{k=1}^{l_{\mathrm{sublayers}}}{\beta}_{kr}{W}_{kr}\ \forall\ k = 1, .\ . \ .\ , l_\mathrm{{sublayers}},  
\end{equation}
$r=l$ is the initial layer from which the penalty is imposed, $k$ is the number of sub-layers,
$l_{\mathrm{sublayers}}$ is the number of sub-layer in each layer block and $\beta$ is one of the profile functions of (\ref{eq:linear_prof}) and (\ref{eq:exp_prof}).\par
In the proposed loss function of (\ref{eq:regularization_func}), we opt for the logarithm function in order to: \textit{(i)} avoid the $\bm{\beta}$ profile functions converging into the constant unity function and \textit{(ii)} prevent gradient explosion, i.e., having the logarithmic decay act as a regularizer \cite{glorot2010understanding}. {Unlike $L_{1}$ norm, the \textit{`log-regularizer'} is non-convex. More broadly, the objective function of an NN is only convex when there are no hidden units, all activations are linear and the design matrix is of full-rank, otherwise, in most cases, the optimization objective is non-convex \cite{Haeffele_2017_CVPR}. To avoid the challenge of having to design an appropriate transformation \cite{malioutov2014iterative}, it is now standard practice to train such NN designs with the combination of stochastic gradient descent (SGD) and appropriate hyper-parameter tuning.} Together with the use of the $L_{1}$ norm, these two aspects enforce sparsity in the network weights corresponding to the lowest part of the $\bm{\beta}$ profile functions when the regularization parameter ($\lambda$) is adequately large \cite{wan2013regularization}. In this way, the model size can be scaled down by expunging some layers deterministically during inference, which reduces memory and computational requirements during model deployment with graceful degradation in detection accuracy.

\subsection{WeSNet with Learnable Weight Profile Coefficients \normalfont{(L-WeSNet)}}
To improve the robustness of the WeSNet against vanishing gradients and possible gradient explosion, the weight profile functions themselves are made trainable parameters, whose values are optimized during  the network training process. This allows for significantly wider exploration of appropriate scaling functions than the predetermined profile functions presented earlier, albeit at the expense of computational complexity during training. To achieve this, (\ref{eq:trainable_params}) is modified to include profile weight functions as learned parameters. 

\begin{equation}
\mathbf{\Tilde{\Phi}} =\{\mathbf{W}_{1r},\ \mathbf{W}_{2r} \ ,\ \mathbf{W}_{3r},\ \bm{b}_{1r},\ \bm{b}_{2r},\ \bm{b}_{3r},\ \bm{\beta}_{r}\}^{L}_{r=1}
\end{equation}
\vspace{2mm}
It is important to note that the monotononicity during training and gradient update is maintained by the shape of the functions of (\ref{eq:linear_prof}) and (\ref{eq:exp_prof}).

{\section{Complexity Analysis}}
WeSNet is a truncated version of DetNet, and the detection is performed at the inference layer (see Fig. \ref{fig:WSVC_model}) by feed forward computation and subsequent application of the soft sign activation function. The computational cost of WeSNet inference is derived based on the cost of operations of an MLP (please see  Appendix \ref{appendix:mlp_complexity} for the details). Our proposed model has 90 layers formed by stacking block of layers, each consisting of three layers DNN. The propagation error is found by computing the derivative of the cost function with respect to the parameters in each block. The computational complexity is specifically measured by the number of operations based on the detector's model. {S}uppose $\mathbf{A}\in \mathbb{C}^{M \times N}$ and $\mathbf{B}\in \mathbb{C}^{N \times L}$ are arbitrary matrices. $\mathbf{D}\in \mathbb{C}^{M \times N}$ is a diagonal matrix, $\mathbf{a,\ b}\in \mathbb{C}^{N \times1}$ and $\mathbf{c}\in \mathbb{C}^{M\times1}$ are arbitrary vectors and $\mathbf{Q}\in \mathbb{C}^{N \times N}$ is positive definite. The required number of FLOPs operations of the standard algebraic expressions of interest to this work are summarized in Table \ref{tab:matrix computational cost}. 

\begin{table*}[hbt!]
\caption{Matrix-vector floating point operations \cite{golub2012matrix}}.
\renewcommand{\arraystretch}{1.3}
\label{tab:matrix computational cost}
    \centering
    \begin{tabular}{c|c|c|c|c}
    \hline
    Expression  &  Description  &   Multiplications &   Summations  &   Total Flops \\
    \hline
    \hline
    $\alpha \mathbf{a}$ &   \text{Vector Scaling}    &   $N$    &   &   $N$ \\
    \hline
    $\alpha \mathbf{A}$ &   \text{Matrix Scaling}   &   $MN$    &   &   $MN$ \\
    \hline
    $ \mathbf{Ab}$ &   \text{Matrix-Vector Prod.}   &   $MN$    &   $M(N-1)$    &   $2MN-M$ \\
    \hline
    $ \mathbf{AB}$ &   \text{Matrix-Matrix Prod.}   &   $MNL$    &   $ML(N-1)$    &   $2MNL-ML$ \\
    \hline
    $ \mathbf{AD}$ &   \text{Matrix-Diagonal Prod.}   &   $MN$    & &   $MN$ \\
    \hline
    $ \mathbf{a}^{H}\mathbf{b}$ &   \text{Inner Prod.}   &   $N$    &   $N-1$    &   $2N-1$ \\
    \hline
    $ \mathbf{a}\mathbf{c}^{H}$ &   \text{Outer Prod.}   &   $MN$    &  & $MN$ \\
    \hline
    $ \mathbf{A}^{H}\mathbf{A}$ &   \text{Gram}   &   $\frac{MN(N+1)}{2}$    &   $\frac{N(M-1)(N+1)}{2}$    &   $MN^{2}+N(M-\frac{N}{2})-\frac{N}{2}$ \\
    \hline
    $ \norm{\mathbf{A}}^{2}$ &   \text{Euclidean norm}   &   $MN$    &  $MN-1$    &   $2MN-1$ \\
    \hline
    $ \mathbf{Q}^{-1}$ &   \text{Inverse of Pos. Definite}   &   $\frac{N^{3}}{2}+\frac{3{N}^{2}}{2}$    &   $\frac{N^{3}}{2}-\frac{N^{2}}{2}$    &   $N^{3}+N^{2}+N$ Including N roots \\
    \hline
    \end{tabular}
\end{table*}

\begin{table*}[hbt!]
\renewcommand{\arraystretch}{1.3}
\caption{MIMO detectors' complexity per symbol slot time.}
\label{tab:complexity-table}
\centering
\begin{tabular}{c|c}
    \hline
    MIMO Detector  &  Number of Flops Operation\\
    \hline
    \hline
    ZF  &   $\left(\frac{56}{3}\right) N^{3}_{t}+38N^{2}_{t}+\left(\frac{28}{3}\right)N_{t}$\\
    \hline
    MMSE   &  $\left(\frac{56}{3}\right) N^{3}_{t}+40N^{2}_{t} +\left(\frac{34}{3}\right) N_{t}+1$ \\
    \hline
    ML  &  $|\mathbf{S} |^{N_{t}}(8N^{2}_{t} +8N_{t}-2)$ \\
    \hline
    SDR  &  $\left(13N^{3}_{t} +25N^{2}_{t}+17N_{t}+4\right)N_{\mathrm{iterations}}$ \cite{ma2004semidefinite, ma2008some}\\
    \hline
    WeSNet    &  $\left[ \left(\tilde{\beta}_{cr}{N}_{t}(128N_{t}+5)+9N_{t}\right)\right]L$, $L= \text{number of layers}$\\
     \hline
    DetNet    &  $\left[\left(N_{t}(128N_{t}-2)\right)\right]L$ \\
    \hline
\end{tabular}

\end{table*}

\begin{table}[!htb]
\renewcommand{\arraystretch}{1.3}
\caption{Simulation settings}
\label{tab:experiment}
\centering
\begin{tabular}{c|c}
    \hline
    Parameters  &  Values\\
    \hline
    \hline
    First Sublayer Dimension  &   $8N_{t}=240$\\
    \hline
    Second Sublayer Dimension   &  $N_{t}=30$ \\
    \hline
    Third Sublayer Dimension  &  $2N_{t}=60$ \\
    \hline
    Number of Layers    &  $L=3N_{t}=90$\\
    \hline
    {Fraction of non-zero Layer Weights}  &  {$\mathbf{\tilde{\beta}}_{\text{cr}}$} \\
    \hline
    Training Samples    &  500000 \\
    \hline
    Batch Size    &  5000 \\
    \hline
    Test Samples    &  50000 \\
    \hline
    Training SNR range    &  8dB - 14dB \\
    \hline
    Test SNR range    &  0dB - 15dB\\ 
    \hline
    Optimizer    &  SGD with Adam \\
    \hline
    Learning Rate    &  0.001 \\
    \hline
    Weight Initializer    &  Xavier Initializer \\
    \hline
    Number of Training Iterations   &  25000 \\
    \hline
    Number of Monte Carlo during inference  &  200 \\
    \hline
\end{tabular}
\end{table}

We use the previous equations and the complexity of the feed-forward inference formulation as detailed in Appendix \ref{appendix:mlp_complexity} to compute the number of floating point operations of each MIMO detector. Our results are summarized in Table  \ref{tab:complexity-table}, and correspond to the following standard assumptions: 

\begin{enumerate}
\item One addition, subtraction of a real number is equal to one computational operation.
\item One multiplication of a complex number is equivalent to four real number multiplications and two real number addition.
\item One addition or subtraction of a complex number is equivalent to two real number additions.
\item One division of a complex number is is equivalent to eight real number multiplications and four additions.\par
\end{enumerate}
Since only a certain fraction of the inputs are used to compute the layer weights of WeSNet and R-WeSNet, most of the operations involved in the feed-forward computations are either sparse vector-matrix multiplication and/or sparse matrix-matrix multiplication. We can evaluate the asymptotic complexity as follows; 
$\bm{\beta}_{1r}\mathbf{W}_{1r}$, $\bm{\beta}_{2r}\mathbf{W}_{2r}$ and $\bm{\beta}_{3r}\mathbf{W}_{3r}$ for detecting a single received symbol are computed by matrix-vector and matrix-matrix multiplications as $\mathcal{O}\left(\sum_{r=1}^{L}\mathbf{u}_{1r}+\mathbf{s}_{2r}+\mathbf{a}_{3r}\right)=\mathcal{O}(n^{3}+n^{2})=\mathcal{O}(n^{3})$.   

\section{Simulation and Numerical Results}\label{section4}
In this section, we present the experimental setup and the performance of the WeSNet under different profile functions and their trainable versions. Amongst deep learning based MIMO detectors, DetNet achieves the best complexity-accuracy performance and also forms the basis of our {proposal}. Therefore, we deploy and benchmark WeSNet against DetNet, but also present performance comparisons against other classical detectors. 
\subsection{Simulation Setup}
WeSNet is implemented in \textit{Tensorflow} 1.12.0 \cite{abadi2016tensorflow} using Python. Since  deep learning libraries only support real number computations, {we use real-valued representation} of the random signals and fading channel to generate the training and test datasets. {The detector is evaluated under both asymmetric (30 transmit and 60 receive antennas) and symmetric channel (16 transmit and 16 receive antennas) conditions.} To ensure a fair comparison with the benchmark model, we use the simulation settings, which are summarized in Table \ref{tab:experiment}. {The detectors we consider are:}

\begin{enumerate}
    \item Linear detectors (\textbf{ZF}\ \text{and}\ \textbf{MMSE}) implemented based on \cite{hung2010improved}.
    \item The optimal detector (\textbf{ML}) and optimization based detector (\textbf{SDR}) based on relaxed semidefinite programming as proposed in \cite{wiesel2005semidefinite} and \cite{ma2008some} respectively.  
    \item The deep learning-based MIMO detectors \textbf{DetNet} as proposed by Samuel \textit{et al.} \cite{samuel2017deep}, Samuel \textit{et al.} \cite{samuel2019learning} and \textbf{OAMP-Net} introduced by Hengtao \textit{et al.}\cite{he2018model}.
\end{enumerate}

\begin{figure}[!bt]
    \centering
    \includegraphics[width=3.7in,height=2.7in]{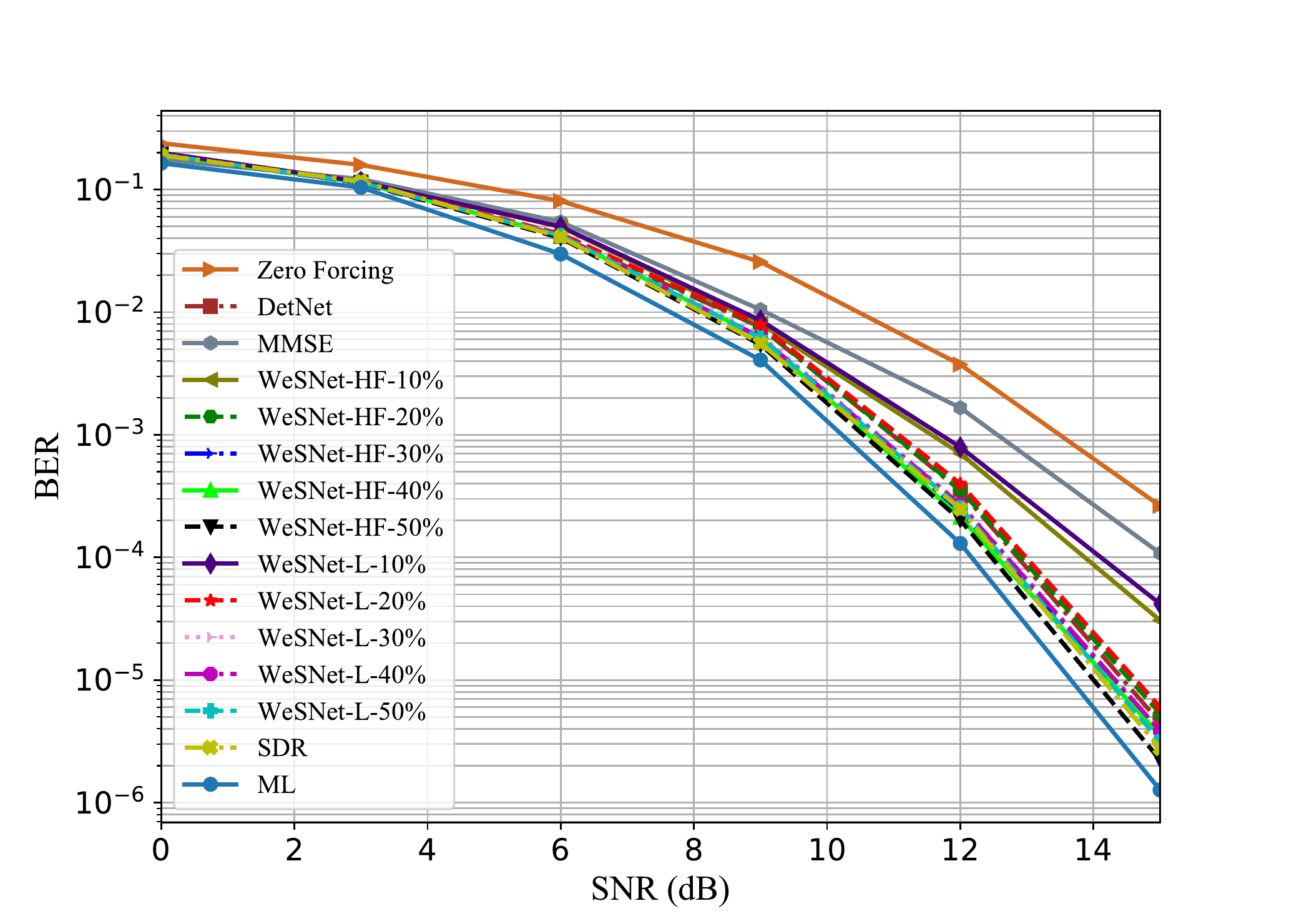}
    \caption{BER comparison of the proposed DNN MIMO Detectors (WeSNet-HF, WeSNet-L), DetNet, ZF, MMSE, SDR and ML under $60 \times30$ fading channel60 using BPSK modulation.}
    \label{fig:ALL_DECT}
\end{figure}

\subsection{Training}
{{First, let us note that training of our model is done once, offline, and can, therefore, accommodate significant complexity followed by the actual deployment of the detector at the inference. Our training dataset comprises transmitted symbols} generated stochastically from random normal distribution drawn from either BPSK or 4-QAM constellation, additive white Gaussian noise (AWGN) generated from a uniform distribution over a wide range of SNR values $\mathcal{U}(\text{8dB - 14dB)}$ and the corresponding received symbols through general random channel taken from a complex Gaussian distribution.} {On the other hand, our inference (test) dataset, is obtained using the same modulation schemes as the training dataset but with different channel instantiations and distinct instantiations of AWGN over over different range of SNR values $\mathcal{U}(\text{0dB - 15dB})$. This training and inference scenario complies with the vast majority of tests in the related literature \cite{samuel2017deep, corlay2018multilevel, Tan2018ImprovingMM, liu2018deep, samuel2019learning, Imanishi2018DeepLI, he2018model, takabe2019trainable}.} We train the model for 25000 iterations with 5000 batch size for each iteration on a standard Intel i7-6700 CPU @ 3.40 GHz processor and use Adam Optimizer\cite{Goodfellow-et-al-2016} for gradient descent optimization. It takes between 17-19 hours to train WeSNet with 20\% and 50\% profile weight coefficients respectively. {This training time is substantial, but it needs to be carried out offline, and only once.} As described in Section II, during training we assume an unknown noise variance and therefore generate the noise vector from a random uniform distribution over the training SNR values $\mathcal{U(} \text{SNR}_{\mathrm{min}} ,\text{SNR}_{\mathrm{max}})$. This allows the network to learn over a wide range of SNR conditions.\par

\begin{figure}[!bt]
    \centering
    \includegraphics[width=3.7in,height=2.7in]{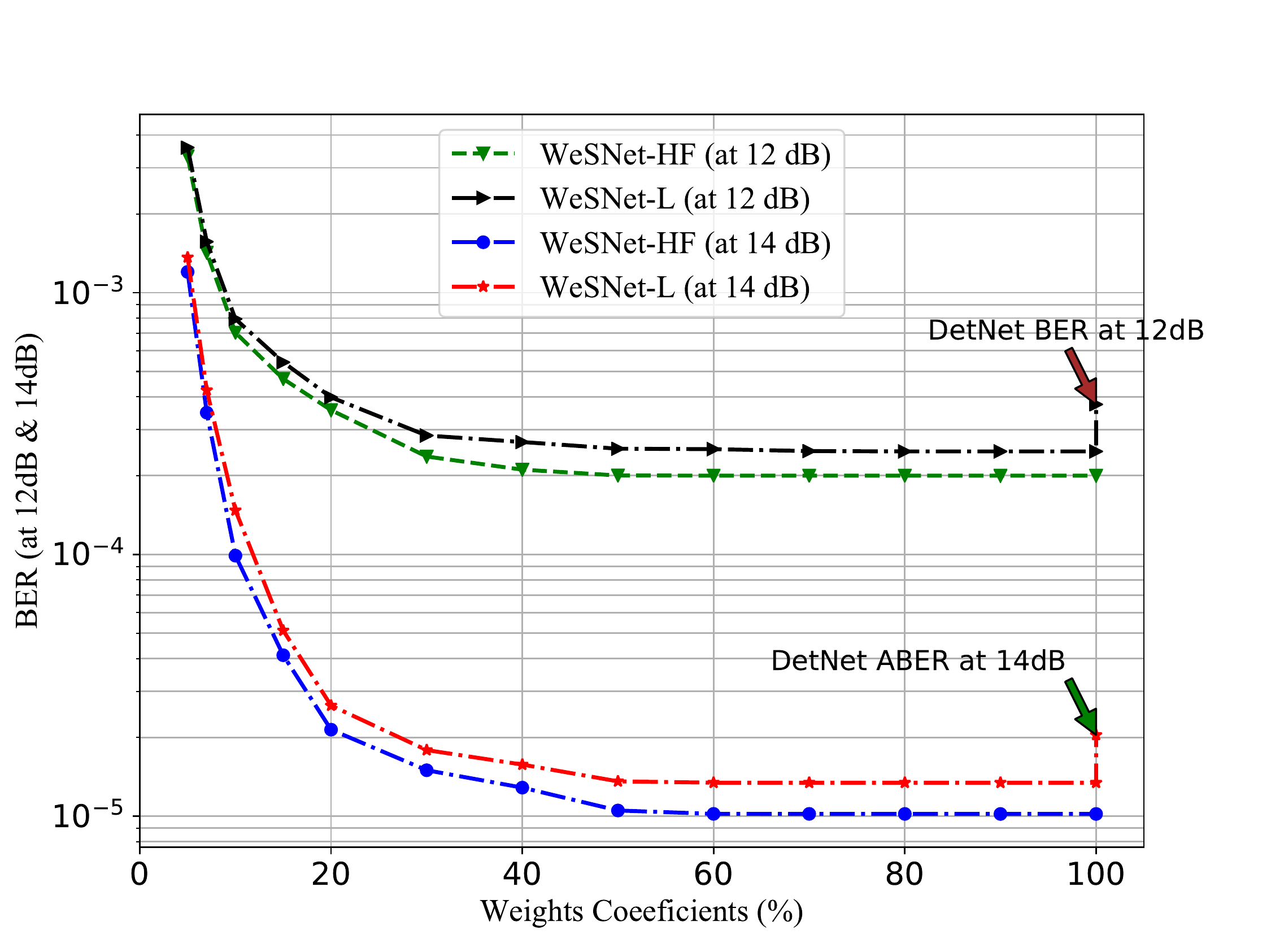}
    \caption{BER vs Percentage Weight Profile Coefficients for WeSNet.}
    \label{fig:avr_ber}
\end{figure}

\subsection{Performance of a WeSNet Realization with Half-Exponential and Linear Profile Functions}
{For clarity, we begin by defining the following term; WeSNet-(HF/L)-x\%: Weight-scaled network obtained from Half-Exponential or Linear profile or function trained and with `x' fraction of the layer weights retained during training and inference.}

Fig. \ref{fig:ALL_DECT} shows the performance of WeSNet with the half-exponential (WeSNet-HF) and linear (WeSNet-L) profile functions of (\ref{eq:linear_prof}) and (\ref{eq:exp_prof}) when retaining increased percentage of inference layers (as marked in the corresponding legends). The benchmarks comprise  DetNet, ZF, MMSE, SDR and ML detectors. Both linear and half-exponential profile WeSNet have comparable performance at lower SNR and profile coefficients between 20\% - 30\% of the layer weights. As expected, the addition of more profile coefficients increases WeSNet's detection accuracy, but performance saturates after 60\% of the coefficients. However, we observe an appreciable difference at higher SNR as more profile weight coefficients are added. At ${10}^{-3}$ BER,  WeSNet can be trained with only 10\% of the layer weights and still outperforms ZF and MMSE by  1.68 dB and 0.79 dB respectively. Overall, WeSNet with only 20\% of the layer weights (WeSNet-HF-20\%) achieves virtually the same performance as our benchmark model (DetNet). In fact, with 50\% profile weight coefficients (WeSNet-HF-50\%), WeSNet outperforms  DetNet, producing the accuracy of symbol detection equivalent to SDR. This gain is an experimental validation that weight profile functions also act as regularizers, i.e., beyond their sparsity-enforcing property, they also avoid overfitting when the model size grows. \par 

{In Fig. \ref{fig:avr_ber},} we show the performance of the WeSNet-HF and WeSNet-L parametric to the utilized layer weight profile coefficients at 12 dB and 14 dB SNRs. Both outperform DetNet and WeSNet-HF surpasses the WeSNet-L at high SNR. For example, at 14 dB and $10^{-5}$ BER, it has gain margin of 0.312 dB over the WeSNet-L. We also observe that the BER at 12 dB and 15 dB SNR improves as more profile coefficients are added, but saturates at 50\% due to weight saturation. This illustrates that, with the addition of profile weight coefficients, at higher SNR the size of the WeStNet can be scaled down during training by almost 40\% - 50\% and still achieve almost identical detection accuracy to the full architecture that retains 100\% of the weights during training.\par

\begin{figure}[!t]
    \centering
    \includegraphics[width=3.7in,height=2.7in]{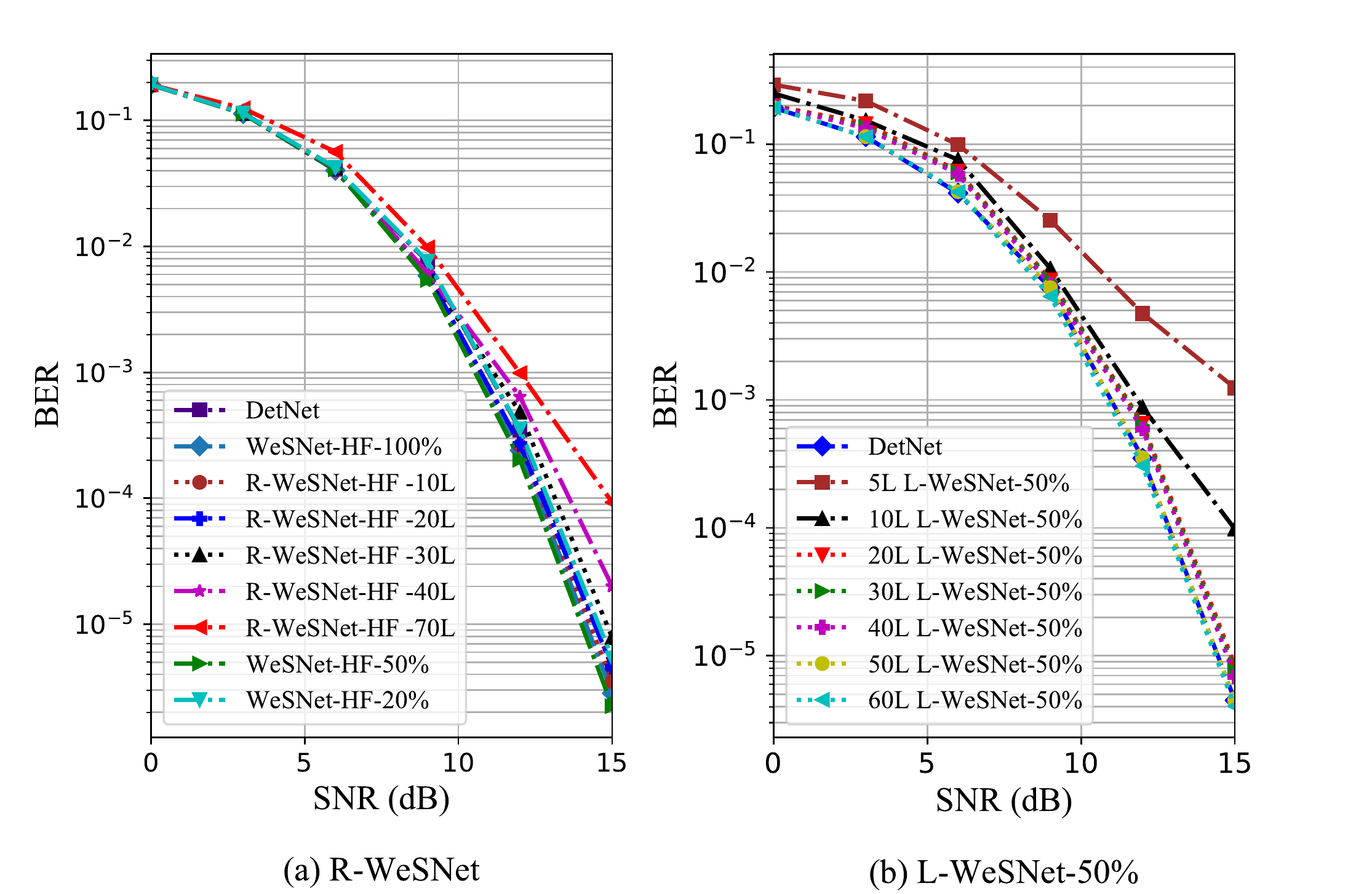}
\caption{Performance comparison of R-WeSNet, L-WeSNet trained with 50\% profile weight coefficients as a function of layers, WeSNet-HF-50\%, WeSNet-HF-20\% and DetNet detectors under $60 \times30$ fading channel60 using BPSK modulation.}
\label{fig:R-WESNET_L-WESNET}
\end{figure}

\begin{figure}[!t]
    \centering
    \includegraphics[width=3.7 in,height=2.7in]{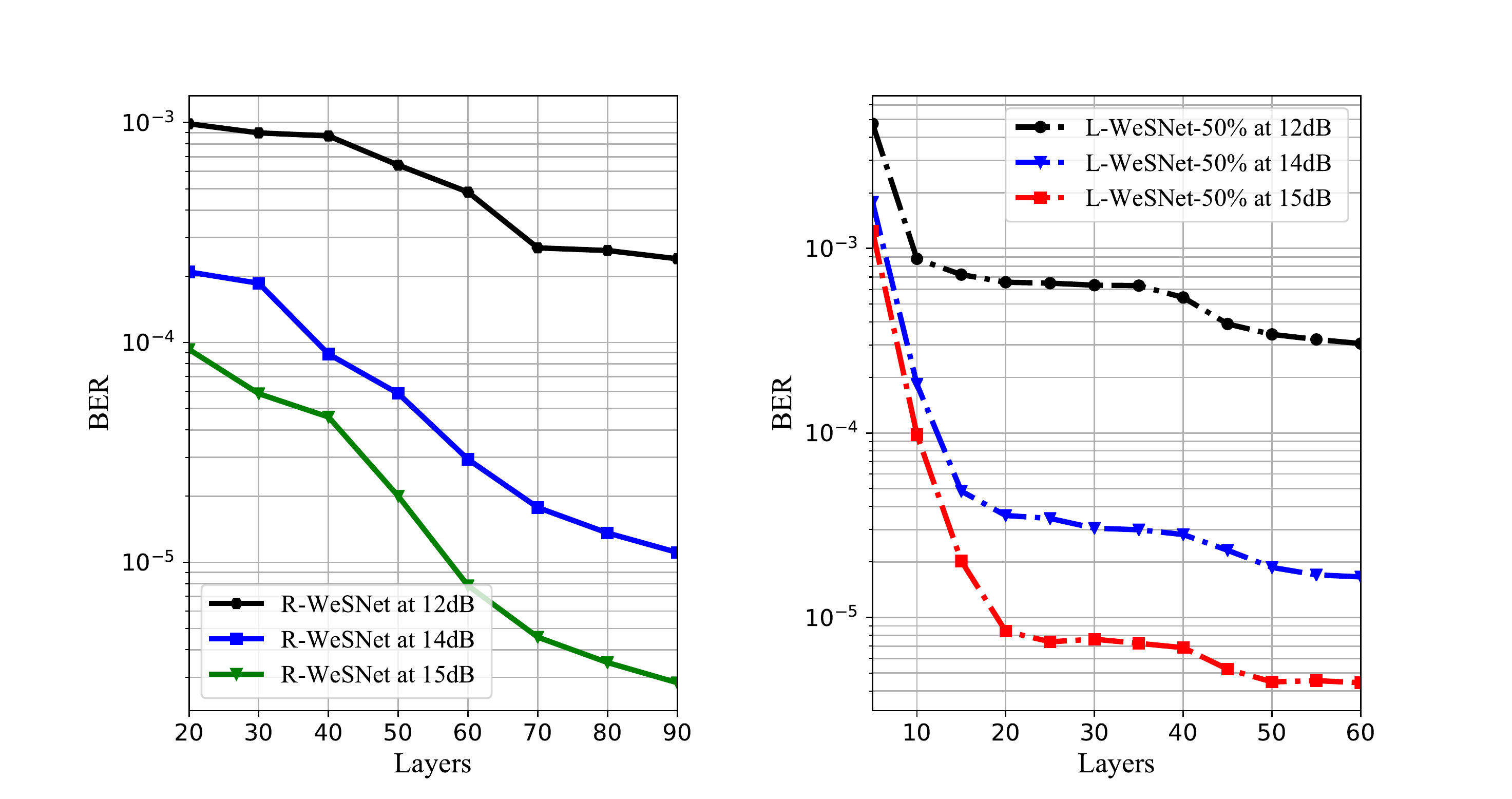}
    \text{\ \ \ \ \ \ \ \ \ \ \footnotesize (a) R-WeSNet.\ \ \ \ \ \ \ \ \ \ \ \ \ \ \ \ \ \ \ (b) L-WeSNet-50\%.}
    \caption{BER for R-WeSNet-HF and L-WeSNet vs number of layers.}
    \label{fig:BER_LAYER_R-WESNET_L-WESNET}
\end{figure}

\subsection{Performance Evaluation of R-WeSNet}
To examine in more detail the performance of our approach against the ``direct'' approach of removing entire layers by enforcing penalty on the weights through \textit{log}-$L_{1}$ regularization, Fig. \ref{fig:R-WESNET_L-WESNET}(a) shows BER-SNR performance curves of the full WeSNet (WeSNet-HF-100\%), DetNet and WeSNet when removing entire layers. The figure shows that removing 70 - 40 layers (R-WeSNet-HF-70L and R-WeSNet-HF-40L) results in considerable loss of accuracy as compared to the corresponding WeSNet-HF models (WeSNet-HF-20\% and WeSNet-HF-50\%). Nevertheless, 20-30 layers (R-WeSNet-HF-20L and R-WeSNet-HF-30L) can be removed while still achieving BER-SNR performance slightly better than the DetNet's. It can be noticed that a R-WeSNet-HF-10L (with 10 layers short-fall) outperforms both WeSNet-HF-50\% and DetNet.\par 

\subsection{BER Performance of L-WeSNet}
In order to examine the performance of our approach when the weight profile coefficients are made learnable (L-WeSNet), Fig. \ref{fig:R-WESNET_L-WESNET}(b) presents the performance of with 50\% learnable weight coefficients (L-WeSNet-HF-50\%) over different number of layers. It can be seen that there is a remarkable performance improvement as the size of the network grows from 5 layers to 60 layers. For instance, at $7.2\times {10}^{-2}$ BER, we observe margin of 2.8 dB between 5 to 30 layers. On the other hand, accuracy remains fairly consistent from 20 to 40 layers. Our study also shows that L-WeSNet-50\% produces the same accuracy as DetNet trained with full 90 layers. This means that, for the studied problem, an efficient deep MIMO detector can be designed with 50\% trainable weight coefficients and 50 layers.\par

\begin{figure}[!t]
    \centering
    \includegraphics[width=3.7in,height=2.7in]{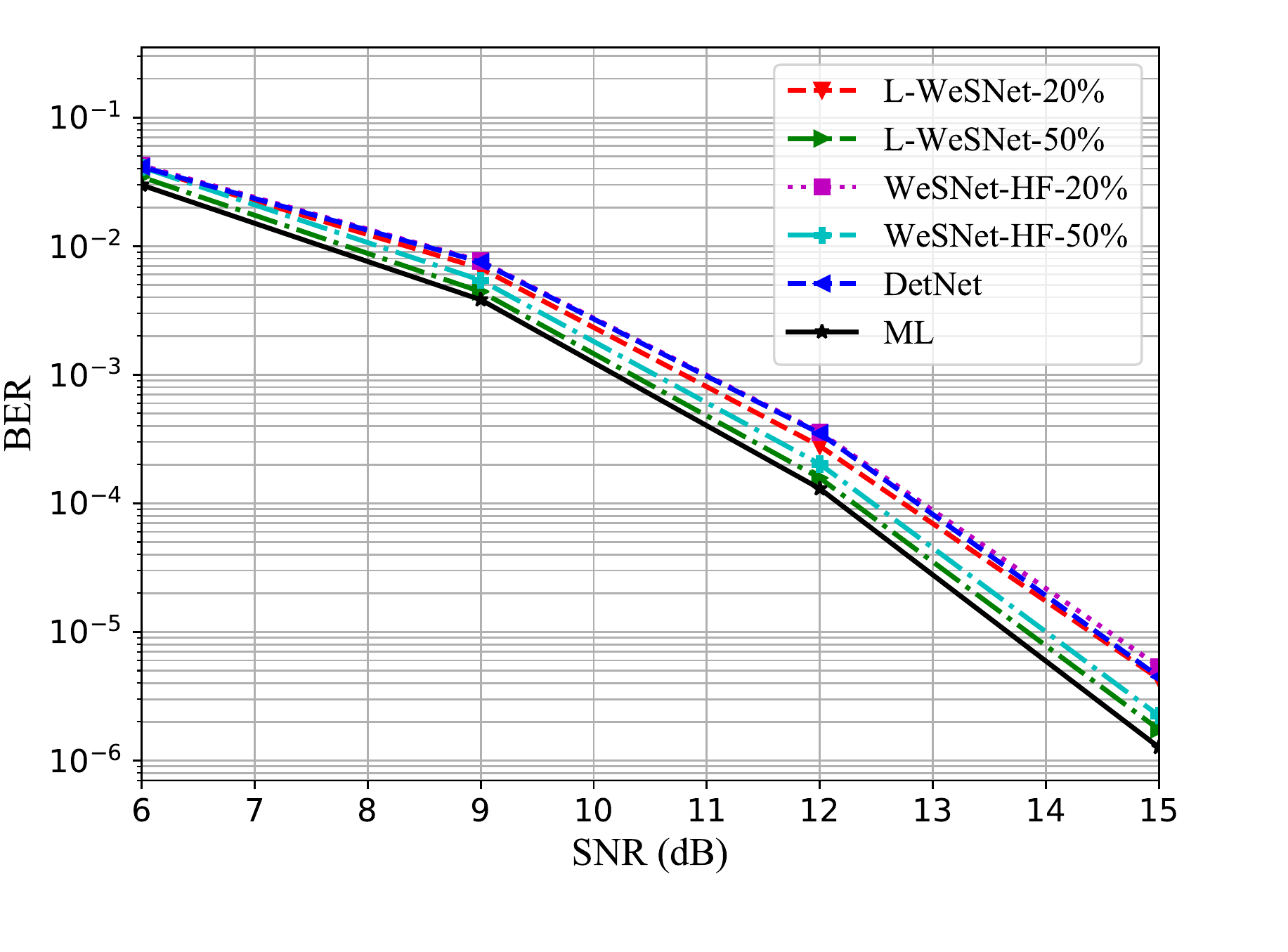}
    \caption{Performance comparison of L-WeSNe, WeSNet, DetNet and ML detectors under $60 \times30$ fading channel using BPSK modulation.}
    \label{fig:BER_L-WESNET_WESNET_ML}
\end{figure}

{Fig. \ref{fig:BER_LAYER_R-WESNET_L-WESNET}} shows the average BER for both R-WeSNet and L-WeSNet against the number of layers at 12 dB, 14 dB and 15 dB SNRs respectively. At 12 dB SNR (Fig. \ref{fig:BER_LAYER_R-WESNET_L-WESNET}(a)), removing 10 - 30 layers during feed forward inference does not significantly change the performance as compared to at 14 dB and 15 dB. However, at 12 dB and 15 dB, we observe a sharp decrease in performance from 70 - 20 layers. The BER is about $10^{-4}$ at 15 dB and less than $10^{-3}$ at 14 dB with 40 layers removed. We also see that, at higher SNR, model size can be reduced significantly by removing up to 50 layers during the inference with slightly loss of accuracy.\par 

No rules or analysis exists to precisely determine the size of a neural network (i.e., number of neurons, layers, or layer parameters) for a specific task. Therefore, we train WeSNet-HF with trainable weight coefficients over the different number of layers to determine the conditions under which we can obtain the minimum BER. The average BER is evaluated at 12 dB, 14 dB and 15 dB for each layer configuration as shown in Fig. \ref{fig:BER_LAYER_R-WESNET_L-WESNET}(b). It can be seen that the BER falls off quickly from 5 - 60 layers. The BER at 14 dB is approximately $3\times10^{-5}$. This value is reasonably constant from 20 - 30 layers and goes down as more layers are added. It can also be seen that at 15 dB, L-WeSNet-50\% produces nearly $10^{-5}$ BER with only 20 layers.\par

\begin{figure*}[!hbt]
\centering
\begin{subfigure}{0.49\textwidth}
\includegraphics[width=3.7in,height=2.7in]{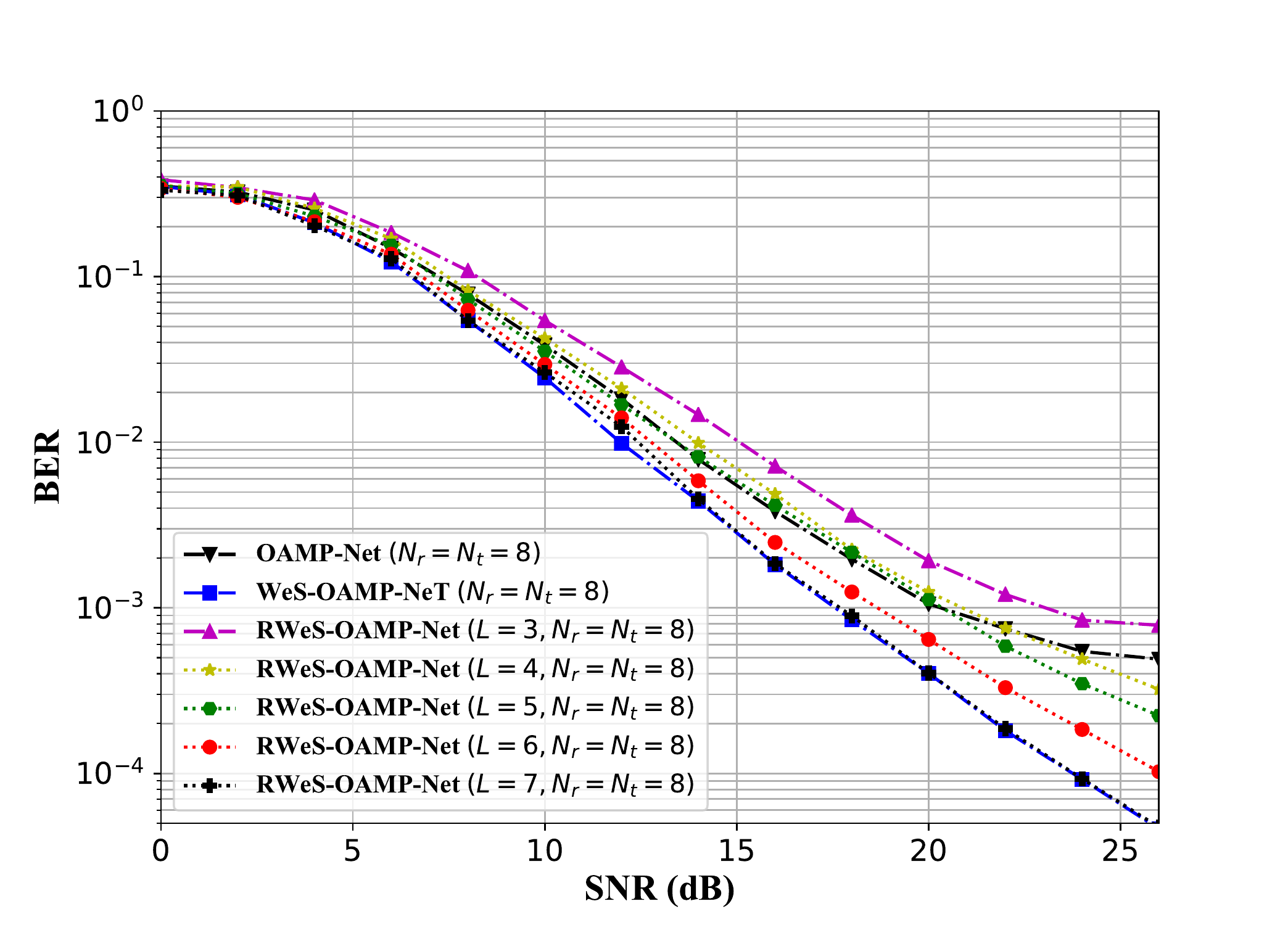}%
\caption{BER vs. SNR over $8 \times8$ fading channel using 4-QAM modulation.}%
\label{fig:WES-OAMP-8X8}%
\end{subfigure}\hfill%
\begin{subfigure}{0.49\textwidth}
\includegraphics[width=3.7in,height=2.7in]{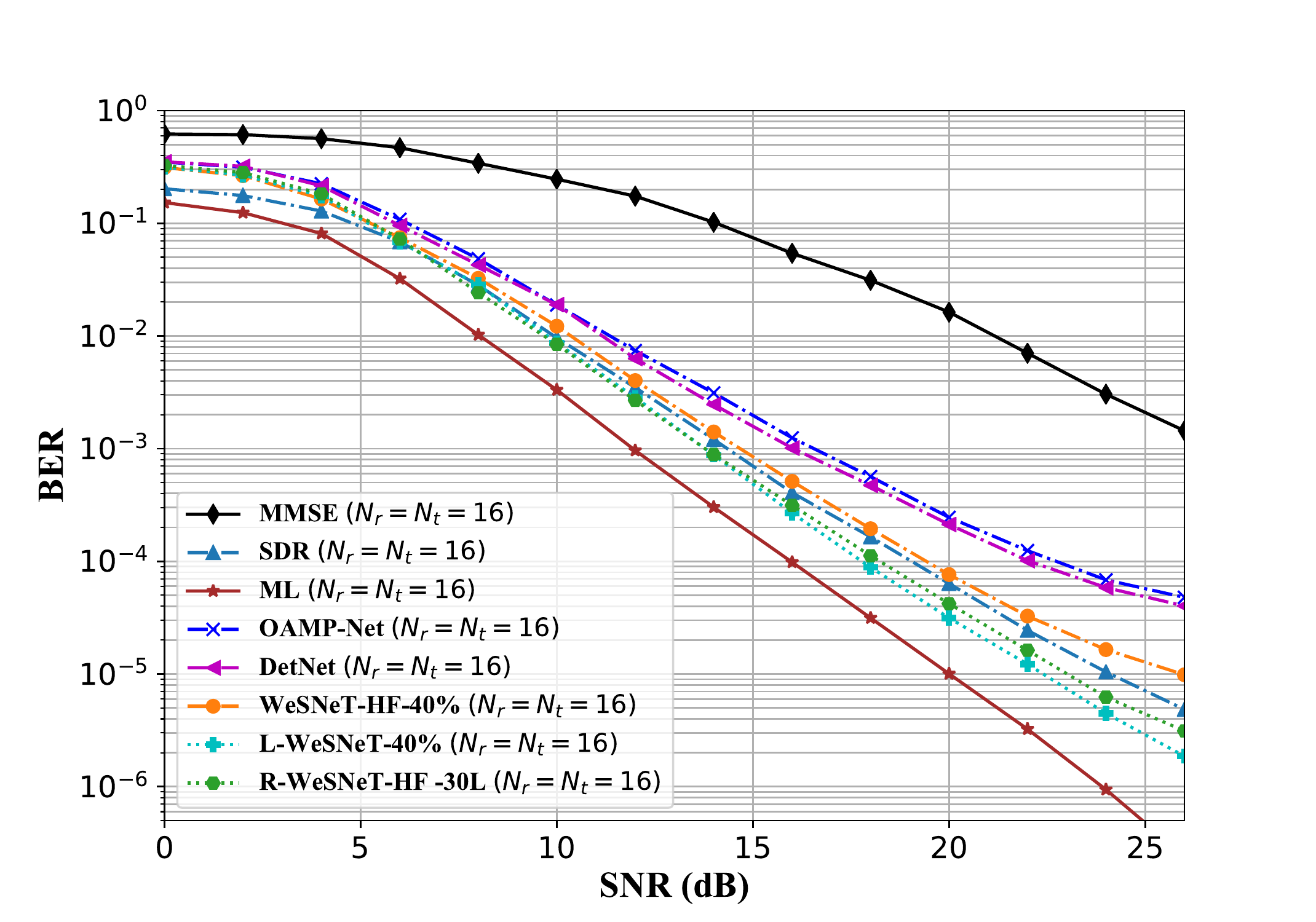}%
\caption{Performance comparison of WeSNet vs benchmark detectors under $16 \times16$ channel using 4-QAM modulation.}%
\label{fig:WES-OAMP-16X16}%
\end{subfigure}\hfill%
\begin{subfigure}{0.49\textwidth}
\includegraphics[width=3.7in,height=2.7in]{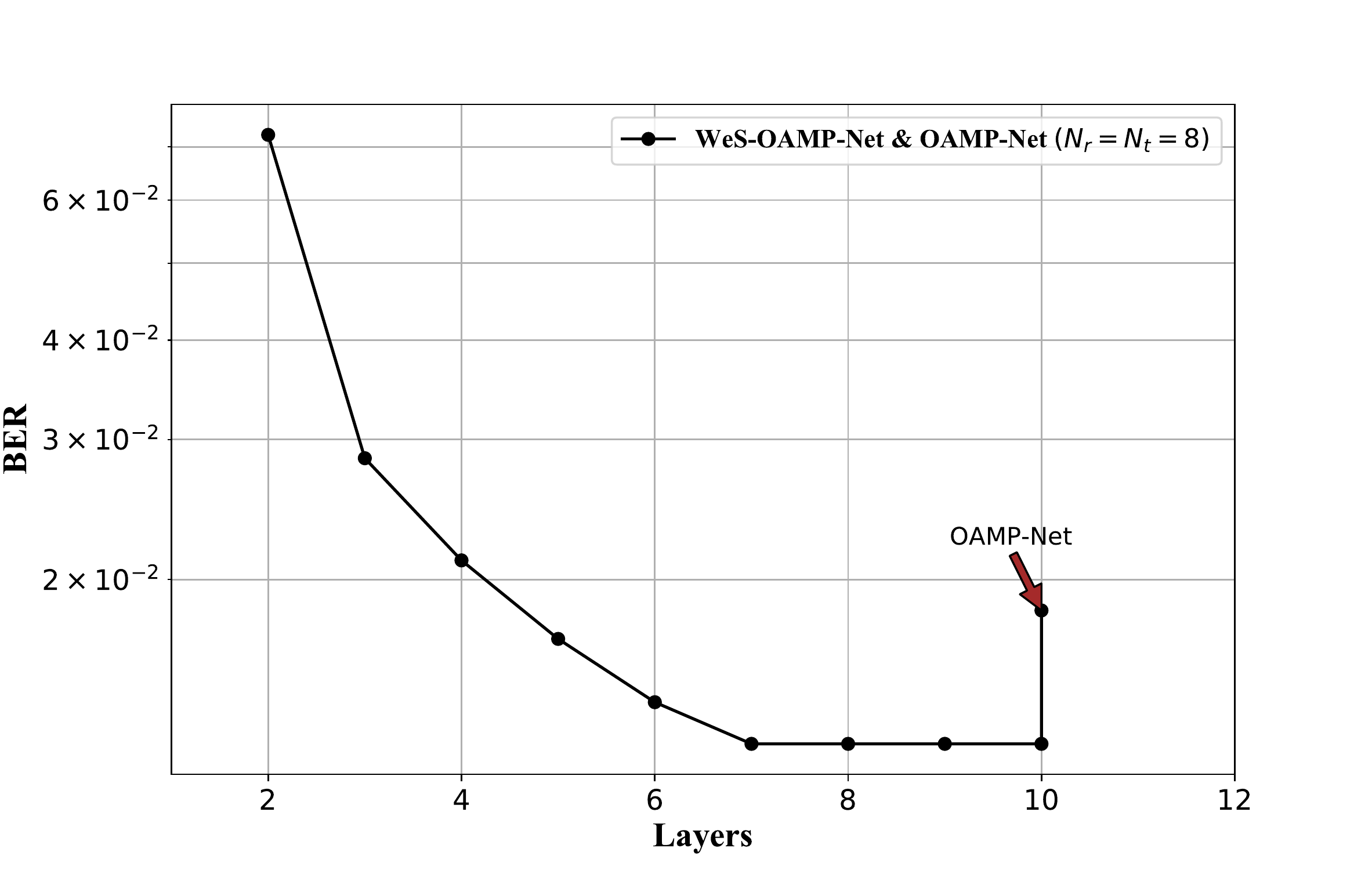}%
\caption{BER vs. Layers over a 8 x 8 fading channel using 4QAM modulation.}%
\label{fig:BER_LAYER-8X8}%
\end{subfigure}%
\caption{Performance evaluation of OAMP-Net vs Weighted-Scaled OAMP-Net (WeS-OAMP-Net).}
\label{fig:PERFORMANCE-M-QAM}
\end{figure*}

In Fig. \ref{fig:BER_L-WESNET_WESNET_ML}, we compare the BER-SNR performance of WeSNet and L-WeSNet both trained over the entire layers with 20\% and 50\% of the profile weight coefficients (L-WeSNet-20\% and L-WeSNet-50\%) against other benchmark models. Our study shows that WeSNet with trainable weight profile functions outperforms the one with non-trainable functions. This comes at the expense of slightly increased training cost due to the additional number of training parameters. This additional training overhead, however, does not increase the inference complexity of the L-WeSNet over WeSNet's, as the inference architectures are the same, except of the difference in the values of the trained weight scaling values. It can be seen that L-WeSNet-20\% at ${10}^{-3} $ BER outperforms both DetNet and WeSNet-HF-20\% by 0.19 dB. Similarly, L-WeSNet-50\% yields better detection accuracy over WeSNet-HF-50\% and DetNet by 0.22 dB and 0.69 dB, respectively.\par

\begin{figure}[!bt]
    \centering\includegraphics[width=3.7in,height=2.7in]{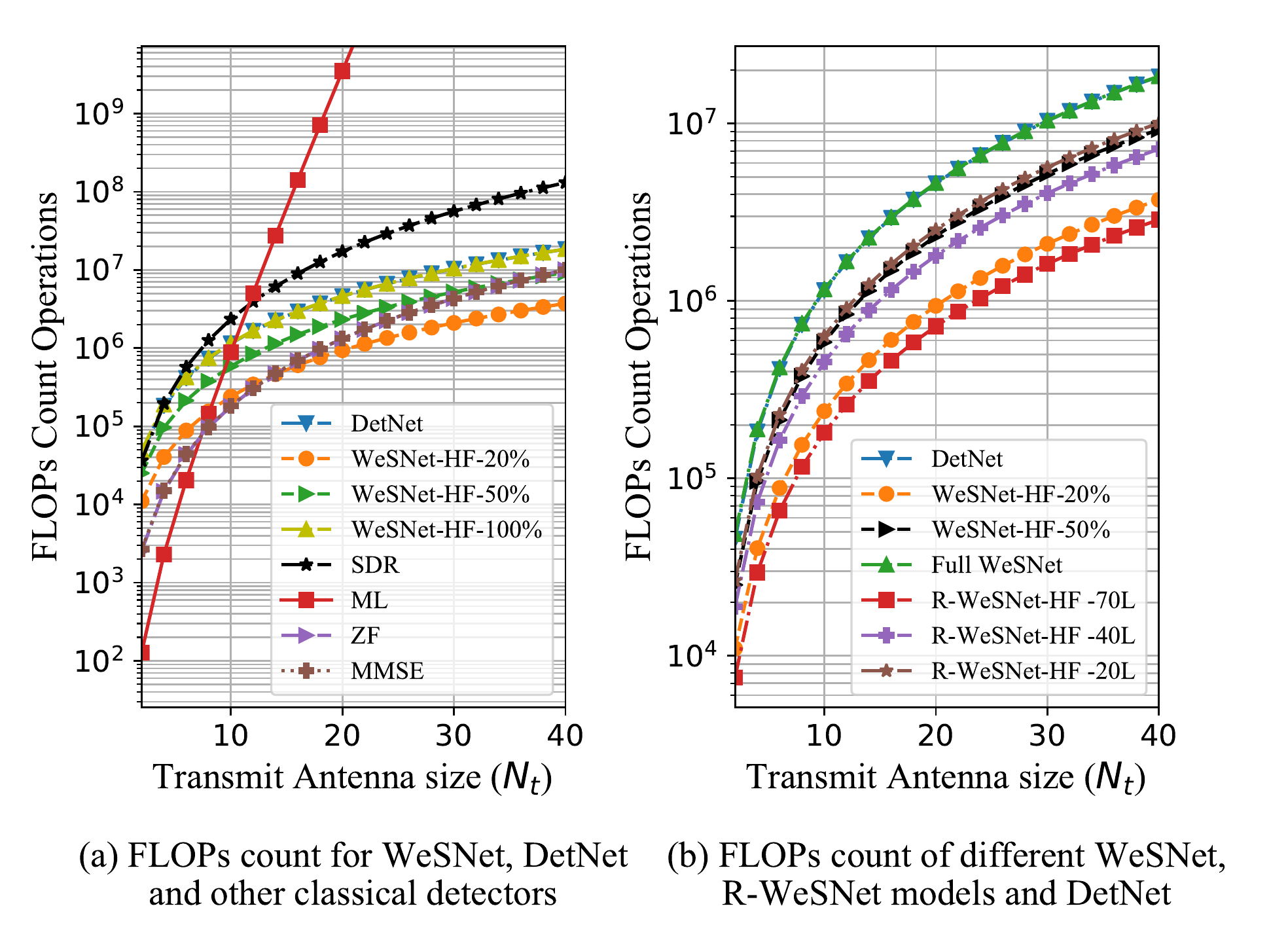}
\caption{Computational complexity comparison of the detectors vs transmit antenna size.}
\label{fig:FLOPS_COUNTS}
\end{figure}

\begin{figure}[!bt]
    \centering
    \includegraphics[width=3.7in,height=2.7in]{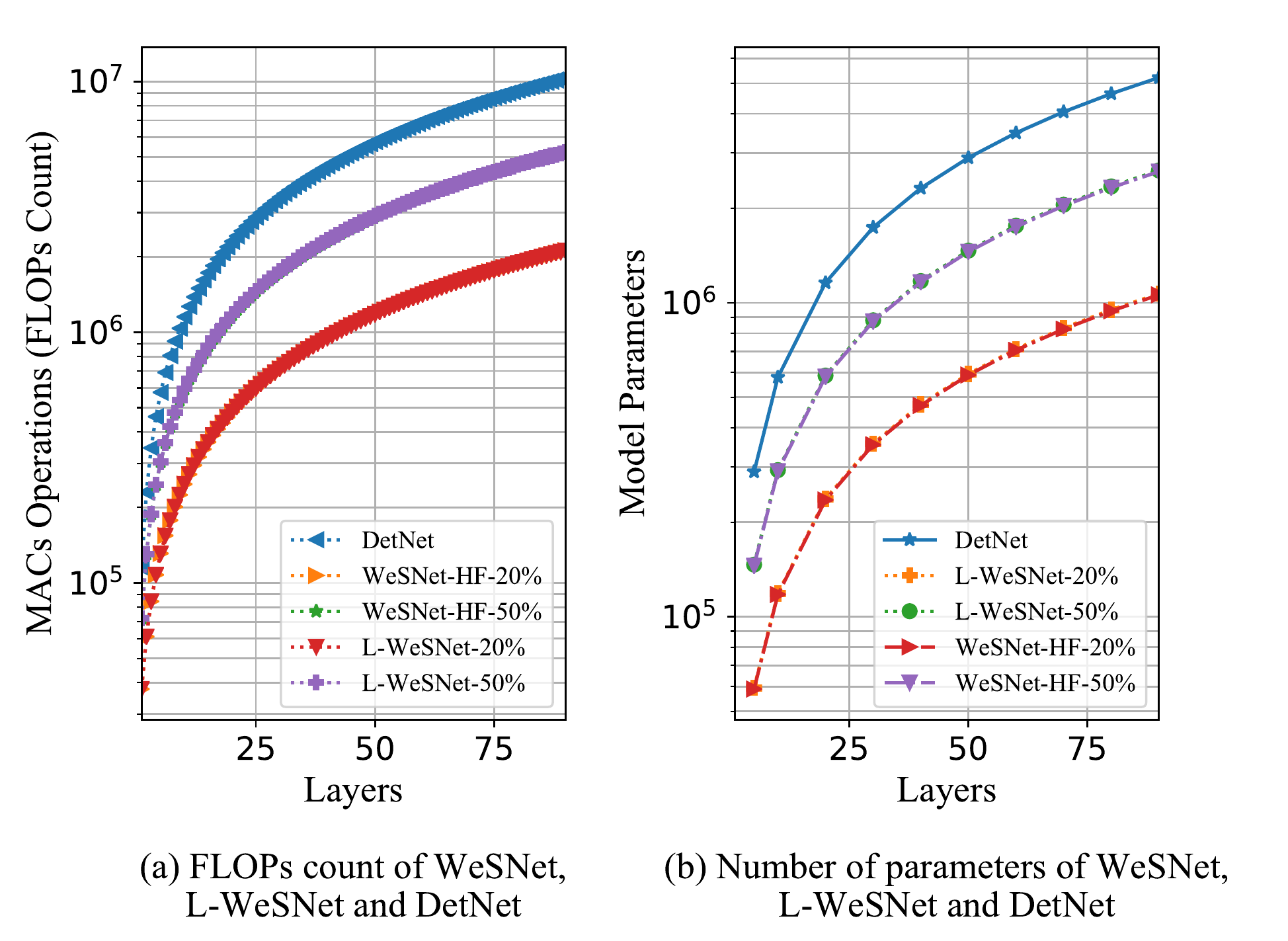}
    \caption{Complexity comparison of WeSNet, L-WeSNet and DetNedt in terms of FLOPs count and model parameters as a function of network layers.}
\label{fig:FLOPS_COUNTS_MODEL_PARAM_LAYER}
\end{figure}

\begin{figure}[!bt]
    \centering
    \includegraphics[width=3.7in,height=2.7in]{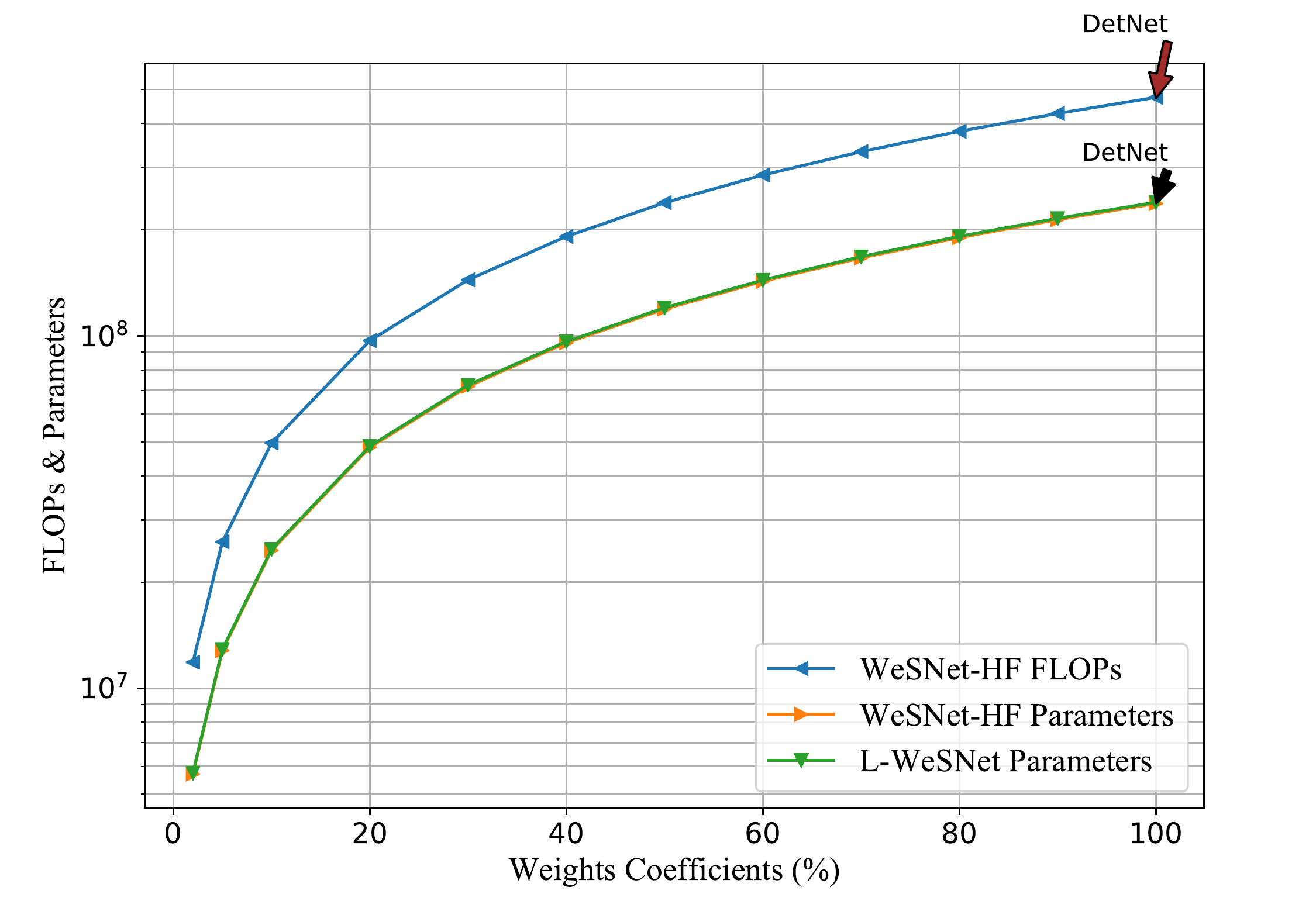}
    \caption{Total FLOPs and model parameters vs weight profile coefficients.}
    \label{fig:FLOPs_Model_params}
\end{figure}

{\subsection{Adaptability of WeSNet beyond the DetNet.}
The proposed approach can be applied to any model that has NN design, including deep unfolding iterative algorithms such as OAMP-Net, TPG-Net, etc. by adding a NN sub-layer design before the estimator and introducing the weight profiling to trade off performance with complexity.} 

{Fig. \ref{fig:WES-OAMP-8X8} shows the performance of the OAMP-Net and its weight-scaled version (Wes-OAMP-Net) designed by introducing the weight-scaling framework to OAMP-Net. In addition, we present results with the regularized Wes-OAMP-Net (RWes-OAMP-Net) under scalable reduction of the utilized layers at inference (from $L=7$ down to $L=3$), showcasing the scalable accuracy-complexity behavior of the proposed framework within the OAMP-Net detector. Wes-OAMP-Net (with 10 layers) and the regularized Wes-OAMP-Net (RWes-OAMP-Net) outperform OAMP-Net (with 10 layers). When scaling down complexity at inference, Wes-OAMP-Net with $L=5$ layers still slightly outperforms conventional OAMP-Net, while the accuracy can be further traded-off for complexity in a graceful manner as more layers are removed during inference.}

{Fig. \ref{fig:WES-OAMP-16X16} shows BER performance of the WeSNet, MMSE, DetNet and OAMP-Net evaluated under symmetric fading channel (16 receive and 16 transmit antennas) using 4-QAM modulation scheme. It can be seen that the performance gap between classical MMSE and ML is significant, i.e., in the range of 15 dB. At lower SNR values (0 - 11 dB), we observe that the performance of WeSNet and OAMP-Net is the same. However, at higher SNR, DetNet outperforms OAMP-Net slightly in the range of 0.5 dB. We also observe that WesNet-HF-40\% and SDR have similar performance, while L-WeSNet-HF-40\% outperforms both of them. Similarly, the regularized WeSNet (R-WeSNet-HF) with 30 layers eliminated during the inference is observed to have outperformed all the receivers. Finally, while the DetNet and OAMP-Net have similar performance, WeSNet outperforms both of them at all SNRs, with a reduced (2 - 3 dB) gap to ML.}\par 
{Finally, the performance-complexity tradeoff is further exemplified in Fig. \ref{fig:BER_LAYER-8X8}, which shows the BER against number of layers. It can be seen that BER decreases as more layers are added and the BER gains saturate at the seventh layer. We also observe that 2.3 - 41.4\% complexity can be saved by reducing the number of OAMP-Net layers from 10 to 3 with loss of accuracy within the range of 0.5 - 3 dB for a system with 8 receive and 8 transmit antennas. Therefore, where before OAMP-Net had only one BER vs complexity operating point in this scenario, our proposed framework has provided a range of BER vs complexity operating points which can be traded-off as per the communication’s link requirements.} 

\subsection{Complexity Evaluation of the Proposed Scheme}
To associate layer sizes with complexity and number of antennas in the MIMO configuration, Fig. \ref{fig:FLOPS_COUNTS}(a) shows the complexity evaluated as the number of FLOPs for WeSNet-HF-100\%, WeSNet-HF-50\%, WeSNet-HF-20\%, DetNet, ZF, MMSE, SDR and ML detectors against the number of transmit antennas. As expected, as the number of antennas increases, the complexity of ML grows exponentially. On the other hand, WeSNet-HF-20\% has the lowest computational cost. As far as the model configuration is concerned, equal number of matrix-matrix and matrix-vector floating point operations are performed by both WeSNet-HF-100\% and DetNet during the feed forward inference. However, WeSNet-HF-50\% and WeSNet-HF-20\% are computationally more efficient than DetNet. For example, with 20\% - 80\% profile weights coefficients, the training of WeSNet-HF is less complex than that of DetNet under the same operating conditions.  When a regularized WeSNet-HF-100\% is trained and layers are removed deterministically at inference, Fig. \ref{fig:FLOPS_COUNTS}(b) shows that the complexity drops, with graceful degradation in performance. Importantly, as expected from prior experiments, the first 30 layers can be abrogated without any significant compromise on the performance.\par 

Fig. \ref{fig:FLOPS_COUNTS_MODEL_PARAM_LAYER}(a) depicts the complexity as function of network layers. The computational requirement grows linearly as more layers are added. It can be observed that the WeSNet-HF-50\% and WeSNet-HF-20\% are less complex than DetNet over the entire range of layers. Our study shows that, at the inference, the complexity of L-WeSNet is not affected by the presence of learnable weight profile functions. Therefore, WeSNet-HF-50\% and WeSNet-HF-20\% and their corresponding learnable versions (L-WeSNet-50\% and L-WeSNet-20\%) have the same computational complexity at inference.\par
Fig. \ref{fig:FLOPS_COUNTS_MODEL_PARAM_LAYER}(b) shows the variation of the model size in terms of number of learnable parameters as a function of network layers. For a given layer dimension (number of neurons), the size of the model is determined by the number of layers and the number of trainable parameters. It can be seen that WeSNet, in addition to having better detection accuracy, it is substantially more memory efficient than DetNet and requires less training time under the same experimental conditions. \par 

As more weights profile coefficients are added, the number of FLOPs increases. Fig. \ref{fig:FLOPs_Model_params} shows how the computational cost and model parameters change with the profile weight coefficients. As shown by earlier experiments, WeSNet-HF achieves performance close to DetNet with only 20\% to 30\% of the layer weights. Therefore, such weight scaling leads to a significant decrease in the model size by 79.82\% and 68.73\% respectively. Similarly, we observe a reduction of 51.43\% computational cost and 49.78\% decrease in model size with 50\% profile weight coefficients.\par 

\section{Conclusion}\label{section5}
In this work, we present an efficient and scalable deep neural network based MIMO detector, where complexity can be adjusted at inference with graceful degradation in the detection accuracy. We introduce a weight scaling framework using monotonically non-increasing profile functions to dynamically prioritize a fraction of the layer weights during training. In order to allow for the neural network architecture to self-adjust to the detection complexity, we also allow for the profile functions themselves to be trainable parameters in the proposed architecture. From our simulation results, we find that the model with trainable coefficients outperforms the one with non-trainable coefficients, but at the cost of   complexity. In addition, our proposal shows that adding  weight scaling via monotonic profile functions maintains the detection accuracy when dropping layer weights. This is achieved in part by introducing an $L_{1}-\text{based}$ regularization function on the layer weights and their profile function coefficients so that the model size can be scaled down by nearly 40\% during the feed-forward inference with marginal impact in the detection accuracy. Source code for training and inference with our proposal is available at: \url{https://github.com/abusaadah/WeSNet\_2019}.

\begin{appendices}
\section{Feed-Forward Computational Cost of an MLP}
\label{appendix:mlp_complexity}
Consider an input, $\mathbf{X} \in \mathbb{R}^{(j,k)}$ and weight $\mathbf{W} \in \mathbb{R}^{(i,j)}$, the linear combination of $\mathbf{X}$ and $\mathbf{W}$ is given by
\begin{equation}
\label{eq:complex1}
\mathbf{{Z}}_{ik}=\mathbf{W}_{ij} \mathbf{X}_{j,k}+\mathbf{b}_{i}
\end{equation}
Applying non linear activation to \eqref{complex1}, gives:
\begin{equation}
\mathbf{a}_{ik}=g(\mathbf{Z}_{ik})
\end{equation}
where $g(\mathord{\cdot})$ is the nonlinear activation function.
The matrix multiplication has an asymptotic computational complexity $\mathcal{O}(n^3)$ and the activation function has  $\mathcal{O}(n)$ complexity.

\subsection{Feed-Forward Inference}
For $N^{[L]}$ number of neurons including bias unit in the ${r}$-{th} layer, the total complexity can be calculated as a sum of the total number of matrix multiplication and the applied activation over the entire layers as
\begin{equation}
    {N}_{\mathrm{matmul}}=\sum_{r=2}^{L}(N^{[r]}N^{[r-1]}N^{[r-2]})+N^{[1]}{N}^{[0]}
\end{equation}
\begin{equation}
   {N}_{g}=\sum_{r=1}^{L}({N})^{[r]}
\end{equation}
\begin{equation}
\begin{split}
\text{Complexity}& = {N}_{\mathrm{matxmul}}+{N}_{g} \\
 & = N_{L} \ .\ N^{3}
\end{split}
\end{equation}
The complexity for ${r}$-{th} layers:
\begin{equation}
\begin{split}
N_{\mathrm{matmul}}& = \mathcal{O}(n \ . \ n^3) \\
 & = \mathcal{O}(n^4)
\end{split}
\end{equation}
Similarly, the complexity ${N}_{g}$ for the activation function with $L$ layers is:
\begin{equation}
\begin{split}
{N_{g}}& = {N_{L}} \ . \ N \\
& = \mathcal{O}(n^2)
\end{split}
\end{equation}
Therefore, the total complexity of the forward propagation is  
\begin{gather}
\label{eq:forward}
\begin{split}
\text{Total complexity}&=\mathcal{O}(n^{4} +n^{2}) \\
&\approx \mathcal{O}(n^{4})
\end{split}
\end{gather}

\end{appendices}

\bibliographystyle{IEEEtran}
\bibliography{ref}

\begin{thebibliography}{10}
\providecommand{\url}[1]{#1}
\csname url@samestyle\endcsname
\providecommand{\newblock}{\relax}
\providecommand{\bibinfo}[2]{#2}
\providecommand{\BIBentrySTDinterwordspacing}{\spaceskip=0pt\relax}
\providecommand{\BIBentryALTinterwordstretchfactor}{4}
\providecommand{\BIBentryALTinterwordspacing}{\spaceskip=\fontdimen2\font plus
\BIBentryALTinterwordstretchfactor\fontdimen3\font minus
  \fontdimen4\font\relax}
\providecommand{\BIBforeignlanguage}[2]{{%
\expandafter\ifx\csname l@#1\endcsname\relax
\typeout{** WARNING: IEEEtran.bst: No hyphenation pattern has been}%
\typeout{** loaded for the language `#1'. Using the pattern for}%
\typeout{** the default language instead.}%
\else
\language=\csname l@#1\endcsname
\fi
#2}}
\providecommand{\BIBdecl}{\relax}
\BIBdecl

\bibitem{wang2014cellular}
C.-X. Wang, F.~Haider, X.~Gao, X.-H. You, Y.~Yang, D.~Yuan, H.~M. Aggoune,
  H.~Haas, S.~Fletcher, and E.~Hepsaydir, ``Cellular architecture and key
  technologies for 5g wireless communication networks,'' \emph{IEEE
  communications magazine}, vol.~52, no.~2, pp. 122--130, 2014.

\bibitem{agiwal2016next}
M.~Agiwal, A.~Roy, and N.~Saxena, ``Next generation 5g wireless networks: A
  comprehensive survey,'' \emph{IEEE Communications Surveys \& Tutorials},
  vol.~18, no.~3, pp. 1617--1655, 2016.

\bibitem{lu2014overview}
L.~Lu, G.~Y. Li, A.~L. Swindlehurst, A.~Ashikhmin, and R.~Zhang, ``An overview
  of massive mimo: Benefits and challenges,'' \emph{IEEE journal of selected
  topics in signal processing}, vol.~8, no.~5, pp. 742--758, 2014.

\bibitem{yao2002lattice}
H.~Yao and G.~W. Wornell, ``Lattice-reduction-aided detectors for mimo
  communication systems,'' in \emph{Global Telecommunications Conference, 2002.
  GLOBECOM'02. IEEE}, vol.~1.\hskip 1em plus 0.5em minus 0.4em\relax IEEE,
  2002, pp. 424--428.

\bibitem{windpassinger2006performance}
C.~Windpassinger, L.~Lampe, R.~F. Fischer, and T.~Hehn, ``A performance study
  of mimo detectors,'' \emph{IEEE Transactions on Wireless Communications},
  vol.~5, no.~8, pp. 2004--2008, 2006.

\bibitem{kailath2005mimo}
T.~Kailath, H.~Vikalo, and B.~Hassibi, ``Mimo receive algorithms,''
  \emph{Space-Time Wireless Systems: From Array Processing to MIMO
  Communications}, vol.~3, p.~2, 2005.

\bibitem{sidiropoulos2006semidefinite}
N.~D. Sidiropoulos and Z.-Q. Luo, ``A semidefinite relaxation approach to mimo
  detection for high-order qam constellations,'' \emph{IEEE signal processing
  letters}, vol.~13, no.~9, pp. 525--528, 2006.

\bibitem{hung2010improved}
C.-Y. Hung and W.-H. Chung, ``An improved mmse-based mimo detection using
  low-complexity constellation search,'' in \emph{2010 IEEE Globecom
  Workshops}.\hskip 1em plus 0.5em minus 0.4em\relax IEEE, 2010, pp. 746--750.

\bibitem{wubben2004near}
D.~Wubben, R.~Bohnke, V.~Kuhn, and K.-D. Kammeyer, ``Near-maximum-likelihood
  detection of mimo systems using mmse-based lattice-reduction,'' in \emph{2004
  IEEE International Conference on Communications (IEEE Cat. No. 04CH37577)},
  vol.~2.\hskip 1em plus 0.5em minus 0.4em\relax IEEE, 2004, pp. 798--802.

\bibitem{zhang2018performance}
C.~Zhang, Y.~Jing, Y.~Huang, and L.~Yang, ``Performance analysis for massive
  mimo downlink with low complexity approximate zero-forcing precoding,''
  \emph{IEEE Transactions on Communications}, vol.~66, no.~9, pp. 3848--3864,
  2018.

\bibitem{Goodfellow-et-al-2016}
I.~Goodfellow, Y.~Bengio, and A.~Courville, \emph{Deep Learning}.\hskip 1em
  plus 0.5em minus 0.4em\relax MIT Press, 2016,
  \url{http://www.deeplearningbook.org}.

\bibitem{simeone2018very}
O.~Simeone, ``A very brief introduction to machine learning with applications
  to communication systems,'' \emph{IEEE Transactions on Cognitive
  Communications and Networking}, vol.~4, no.~4, pp. 648--664, 2018.

\bibitem{qin2019deep}
Z.~Qin, H.~Ye, G.~Y. Li, and B.-H.~F. Juang, ``Deep learning in physical layer
  communications,'' \emph{IEEE Wireless Communications}, vol.~26, no.~2, pp.
  93--99, 2019.

\bibitem{gruber2017deep}
T.~Gruber, S.~Cammerer, J.~Hoydis, and S.~ten Brink, ``On deep learning-based
  channel decoding,'' in \emph{2017 51st Annual Conference on Information
  Sciences and Systems (CISS)}.\hskip 1em plus 0.5em minus 0.4em\relax IEEE,
  2017, pp. 1--6.

\bibitem{nachmani2016learning}
E.~Nachmani, Y.~Be'ery, and D.~Burshtein, ``Learning to decode linear codes
  using deep learning,'' in \emph{2016 54th Annual Allerton Conference on
  Communication, Control, and Computing (Allerton)}.\hskip 1em plus 0.5em minus
  0.4em\relax IEEE, 2016, pp. 341--346.

\bibitem{o2016learning}
T.~J. O'Shea, K.~Karra, and T.~C. Clancy, ``Learning to communicate: Channel
  auto-encoders, domain specific regularizers, and attention,'' in \emph{2016
  IEEE International Symposium on Signal Processing and Information Technology
  (ISSPIT)}.\hskip 1em plus 0.5em minus 0.4em\relax IEEE, 2016, pp. 223--228.

\bibitem{o2017introduction}
T.~O’Shea and J.~Hoydis, ``An introduction to deep learning for the physical
  layer,'' \emph{IEEE Transactions on Cognitive Communications and Networking},
  vol.~3, no.~4, pp. 563--575, 2017.

\bibitem{dorner2018deep}
S.~D{\"o}rner, S.~Cammerer, J.~Hoydis, and S.~ten Brink, ``Deep learning based
  communication over the air,'' \emph{IEEE Journal of Selected Topics in Signal
  Processing}, vol.~12, no.~1, pp. 132--143, 2018.

\bibitem{OShea2017DeepLB}
T.~J. O'Shea, T.~Erpek, and T.~C. Clancy, ``Deep learning based mimo
  communications,'' \emph{CoRR}, vol. abs/1707.07980, 2017.

\bibitem{xu2018joint}
W.~Xu, Z.~Zhong, Y.~Be'ery, X.~You, and C.~Zhang, ``Joint neural network
  equalizer and decoder,'' in \emph{2018 15th International Symposium on
  Wireless Communication Systems (ISWCS)}.\hskip 1em plus 0.5em minus
  0.4em\relax IEEE, 2018, pp. 1--5.

\bibitem{9020494}
A.~{Balatsoukas-Stimming} and C.~{Studer}, ``Deep unfolding for communications
  systems: A survey and some new directions,'' in \emph{2019 IEEE International
  Workshop on Signal Processing Systems (SiPS)}, 2019, pp. 266--271.

\bibitem{samuel2017deep}
N.~Samuel, T.~Diskin, and A.~Wiesel, ``Deep mimo detection,'' in \emph{Signal
  Processing Advances in Wireless Communications (SPAWC), 2017 IEEE 18th
  International Workshop on}.\hskip 1em plus 0.5em minus 0.4em\relax IEEE,
  2017, pp. 1--5.

\bibitem{corlay2018multilevel}
V.~Corlay, J.~J. Boutros, P.~Ciblat, and L.~Brunel, ``Multilevel mimo detection
  with deep learning,'' in \emph{2018 52nd Asilomar Conference on Signals,
  Systems, and Computers}.\hskip 1em plus 0.5em minus 0.4em\relax IEEE, 2018,
  pp. 1805--1809.

\bibitem{Tan2018ImprovingMM}
X.~Tan, W.~Xu, Y.~Be'ery, Z.~Zhang, X.~You, and C.~Zhang, ``Improving massive
  mimo belief propagation detector with deep neural network,'' \emph{CoRR},
  vol. abs/1804.01002, 2018.

\bibitem{liu2018deep}
X.~Liu and Y.~Li, ``Deep mimo detection based on belief propagation,'' in
  \emph{2018 IEEE Information Theory Workshop (ITW)}.\hskip 1em plus 0.5em
  minus 0.4em\relax IEEE, 2018, pp. 1--5.

\bibitem{samuel2019learning}
N.~Samuel, A.~Wiesel, and T.~Diskin, ``Learning to detect,'' \emph{IEEE
  Transactions on Signal Processing}, 2019.

\bibitem{Imanishi2018DeepLI}
M.~Imanishi, S.~Takabe, and T.~Wadayama, ``Deep learning-aided iterative
  detector for massive overloaded mimo channels,'' \emph{CoRR}, vol.
  abs/1806.10827, 2018.

\bibitem{he2018model}
H.~He, C.-K. Wen, S.~Jin, and G.~Y. Li, ``A model-driven deep learning network
  for mimo detection,'' in \emph{2018 IEEE Global Conference on Signal and
  Information Processing (GlobalSIP)}.\hskip 1em plus 0.5em minus 0.4em\relax
  IEEE, 2018, pp. 584--588.

\bibitem{takabe2019trainable}
S.~Takabe, M.~Imanishi, T.~Wadayama, R.~Hayakawa, and K.~Hayashi, ``Trainable
  projected gradient detector for massive overloaded mimo channels: Data-driven
  tuning approach,'' \emph{IEEE Access}, vol.~7, pp. 93\,326--93\,338, 2019.

\bibitem{sze2017efficient}
V.~Sze, Y.-H. Chen, T.-J. Yang, and J.~S. Emer, ``Efficient processing of deep
  neural networks: A tutorial and survey,'' \emph{Proceedings of the IEEE},
  vol. 105, no.~12, pp. 2295--2329, 2017.

\bibitem{he2017channel}
Y.~He, X.~Zhang, and J.~Sun, ``Channel pruning for accelerating very deep
  neural networks,'' in \emph{Proceedings of the IEEE International Conference
  on Computer Vision}, 2017, pp. 1389--1397.

\bibitem{zhu2018adaptive}
X.~Zhu, W.~Zhou, and H.~Li, ``Adaptive layerwise quantization for deep neural
  network compression,'' in \emph{2018 IEEE International Conference on
  Multimedia and Expo (ICME)}.\hskip 1em plus 0.5em minus 0.4em\relax IEEE,
  2018, pp. 1--6.

\bibitem{srivastava2014dropout}
N.~Srivastava, G.~Hinton, A.~Krizhevsky, I.~Sutskever, and R.~Salakhutdinov,
  ``Dropout: a simple way to prevent neural networks from overfitting,''
  \emph{The Journal of Machine Learning Research}, vol.~15, no.~1, pp.
  1929--1958, 2014.

\bibitem{wan2013regularization}
L.~Wan, M.~Zeiler, S.~Zhang, Y.~Le~Cun, and R.~Fergus, ``Regularization of
  neural networks using dropconnect,'' in \emph{International conference on
  machine learning}, 2013, pp. 1058--1066.

\bibitem{rastegari2016xnor}
M.~Rastegari, V.~Ordonez, J.~Redmon, and A.~Farhadi, ``Xnor-net: Imagenet
  classification using binary convolutional neural networks,'' in
  \emph{European Conference on Computer Vision}.\hskip 1em plus 0.5em minus
  0.4em\relax Springer, 2016, pp. 525--542.

\bibitem{hubara2016binarized}
I.~Hubara, M.~Courbariaux, D.~Soudry, R.~El-Yaniv, and Y.~Bengio, ``Binarized
  neural networks,'' in \emph{Advances in neural information processing
  systems}, 2016, pp. 4107--4115.

\bibitem{wang1996artificial}
X.-A. Wang and S.~B. Wicker, ``An artificial neural net viterbi decoder,''
  \emph{IEEE Transactions on communications}, vol.~44, no.~2, pp. 165--171,
  1996.

\bibitem{mcdanel2017incomplete}
B.~McDanel, S.~Teerapittayanon, and H.~Kung, ``Incomplete dot products for
  dynamic computation scaling in neural network inference,'' in \emph{2017 16th
  IEEE International Conference on Machine Learning and Applications
  (ICMLA)}.\hskip 1em plus 0.5em minus 0.4em\relax IEEE, 2017, pp. 186--193.

\bibitem{ma2004semidefinite}
W.-K. Ma, P.-C. Ching, and Z.~Ding, ``Semidefinite relaxation based multiuser
  detection for m-ary psk multiuser systems,'' \emph{IEEE Transactions on
  Signal Processing}, vol.~52, no.~10, pp. 2862--2872, 2004.

\bibitem{5447068}
Z.~{Luo}, W.~{Ma}, A.~M. {So}, Y.~{Ye}, and S.~{Zhang}, ``Semidefinite
  relaxation of quadratic optimization problems,'' \emph{IEEE Signal Processing
  Magazine}, vol.~27, no.~3, pp. 20--34, May 2010.

\bibitem{wiesel2005semidefinite}
A.~Wiesel, Y.~C. Eldar, and S.~Shamai, ``Semidefinite relaxation for detection
  of 16-qam signaling in mimo channels,'' \emph{IEEE Signal Processing
  Letters}, vol.~12, no.~9, pp. 653--656, 2005.

\bibitem{scardapane2017group}
S.~Scardapane, D.~Comminiello, A.~Hussain, and A.~Uncini, ``Group sparse
  regularization for deep neural networks,'' \emph{Neurocomputing}, vol. 241,
  pp. 81--89, 2017.

\bibitem{glorot2010understanding}
X.~Glorot and Y.~Bengio, ``Understanding the difficulty of training deep
  feedforward neural networks,'' in \emph{Proceedings of the thirteenth
  international conference on artificial intelligence and statistics}, 2010,
  pp. 249--256.

\bibitem{Haeffele_2017_CVPR}
B.~D. Haeffele and R.~Vidal, ``Global optimality in neural network training,''
  in \emph{The IEEE Conference on Computer Vision and Pattern Recognition
  (CVPR)}, July 2017.

\bibitem{malioutov2014iterative}
D.~Malioutov and A.~Aravkin, ``Iterative log thresholding,'' in \emph{2014 IEEE
  International Conference on Acoustics, Speech and Signal Processing
  (ICASSP)}.\hskip 1em plus 0.5em minus 0.4em\relax IEEE, 2014, pp. 7198--7202.

\bibitem{golub2012matrix}
G.~H. Golub and C.~F. Van~Loan, \emph{Matrix computations}.\hskip 1em plus
  0.5em minus 0.4em\relax JHU press, 2012, vol.~3.

\bibitem{ma2008some}
W.-K. Ma, C.-C. Su, J.~Jald{\'e}n, and C.-Y. Chi, ``Some results on 16-qam mimo
  detection using semidefinite relaxation,'' in \emph{2008 IEEE International
  Conference on Acoustics, Speech and Signal Processing}.\hskip 1em plus 0.5em
  minus 0.4em\relax IEEE, 2008, pp. 2673--2676.

\bibitem{abadi2016tensorflow}
M.~Abadi, P.~Barham, J.~Chen, Z.~Chen, A.~Davis, J.~Dean, M.~Devin,
  S.~Ghemawat, G.~Irving, M.~Isard \emph{et~al.}, ``Tensorflow: A system for
  large-scale machine learning,'' in \emph{12th $\{$USENIX$\}$ Symposium on
  Operating Systems Design and Implementation ($\{$OSDI$\}$ 16)}, 2016, pp.
  265--283.

\end{thebibliography}

\vspace{-19mm}


\end{document}